%% file: lambda-delta.tex
\documentclass[acmtocl]{acmtrans2m}
\newtheorem{theorem}{Theorem}
\newtheorem{definition}{Definition}

\usepackage{macros}

\newcommand\titletext{The Formal System $\LD$}

\author{
 FERRUCCIO GUIDI \\ 
 Department of Computer Science, University of Bologna, Italy
}

\title{\titletext}

\begin{abstract} 
\input{abstract}
\end{abstract}

            
            

\begin{document}

\maketitle

\begin{bottomstuff}
\end{bottomstuff}

\input{dedication}

\input{introduction}
\input{motivations}
\input{outline}
\input{specification}

\input{definitions}
\input{language}

\input{helpers}
\input{reduction_defs}
\input{native_defs}
\input{static_defs}

\input{arity_defs}
\input{preorders}

\input{theory}
\input{arity_props}
\input{reduction_props}
\input{native_props}
\input{static_props}
\input{examples}

\input{extension}
\input{exclusion}
\input{legal}

\input{conclusions}
\input{blocks}
\input{todo}

\appendix

\input{mtt}

\input{expressions}
\input{judgements}
\input{rules}

\input{duality}
\input{complete}
\input{aggregates}
\input{polarity}

\input{progress}

\input{proofs}

\begin{acks}
\input{thanks}
\end{acks}

\bibliography{references,lambda-delta,kamareddine,spa,helm,pml}

\begin{received}
Received November 2006;
revised May 2008;
accepted July 2008
\end{received}

\end{document}

%% file: abstract.tex
The formal system $\LD$ is a typed $\lambda$-calculus that pursues the
unification of terms, types, environments and contexts as the main goal.
$\LD$ takes some features from the Automath-related $\lambda$-calculi and some
from the pure type systems, but differs from both in that it does not
include the $\Pi$ construction while it provides for an abbreviation mechanism
at the level of terms.
$\LD$ enjoys some important desirable properties such as the confluence of
reduction, the correctness of types, the uniqueness of types up to conversion,
the subject reduction of the type assignment, the strong normalization of
the typed terms and, as a corollary, the decidability of type inference
problem.

%% file: dedication.tex
\begin{flushright}\begin{footnotesize}
\begin{tabular}{c}
To Silvia,\\a very special lady
\end{tabular}
\end{footnotesize}\end{flushright}

%% file: introduction.tex
\section{Introduction}
\seclabel{intro}

The leading goal at the root of the present work is the design of a typed
$\lambda$-calculus, to be used as a logical framework, featuring the
unification of terms, types and environments (with the terminology of
\cite{SU06}) while enjoying a desirable meta-theory in the sense of
\cite{Brn92}.
In principle we pursue this unification, whose benefits we discuss in
\subsecref{motivations}, by defining a suitable set of expressions that can
be terms, types and environments at the same time.

The purpose of this paper is to report on our first attempt to realize such a
calculus.
In \subsecref{outline} we summarize our starting points and our achievements.

In \subsecref{specification} we briefly introduce the digital specification of
our calculus and of its theory inside the $\CIC$ (CIC) \cite{Gui05}.
This specification has been checked by two CIC-based proof assistants.

The calculus is defined in \secref{definitions} where the syntax, the reduction
rules and the type assignment rules are given.
Our main theorems on the calculus are presented in \secref{theory}.
In \secref{extension} we extend our calculus by adding an ``exclusion''
binder, which we show an application of.
The concluding remarks are in \secref{conclusions}.

This paper includes four appendices: in \appref{mtt} we show an application of
our calculus as a theory of expressions for the structural fragment of the
Minimal Type Theory \cite{MS05}, while in \appref{duality} the author proposes
to push  the calculus in the direction of the ``environments as terms as
types'' paradigm until the unification of these three concepts is reached.

In \appref{progress} we report on the differences between the version of the
calculus in front of the reader and its initial version \cite{Gui06a}.

In \appref{proofs} we give the pointers to the digital version of our results.

\xcomment{

}

%% file: motivations.tex
\subsection{Background and Motivations}
\subseclabel{motivations}

Untyped $\lambda$-calculus \cite{Chu41} was introduced by Church as a theory
of computable functions. 
Adding a very simple type theory to this calculus, where types are
never created by abstraction, Curry obtained a version of the simply typed
$\lambda$-calculus $\LR$ (a different version of $\LR$ was proposed by Church
afterwords).

Typing by abstraction was introduced in the second half of the past century in
response to the need of improving the expressiveness of the above type theory,
and this gave rise to many $\lambda$-calculi typed more powerfully.
The type of a term is always assigned in an environment, that is a structure
holding the type information on the free variables that may occur in that term
\cite{SU06}.

An historical survey on type theory can be found in \cite{KLN04}.

In some theories a type can be treated as a term and can be given a type,
which is usually termed a kind.
Nevertheless many calculi, especially those of the Pure Type Systems (PTS)
tradition \cite{Brn92}, provide for constructions that build types, or kinds,
but not terms. 
This is the case of the so-called $\Pi$ construction.
Moreover terms and environments usually belong to distinct syntactical
categories.

One reason for having different constructions for terms and types lays in
the so-called ``Propositions As Types and Proof As Terms'' (PAT)
interpretation \cite{KLN04} (also known as the Curry-Howard isomorphism)
and in the general consensus that propositions and proofs have a
significantly different structure. 
We recall that according to the PAT interpretation, a typed $\lambda$-calculus
can serve as a logical framework where a proposition is encoded in a type
whose inhabitants encode the proofs of that proposition.

On the other hand there are scenarios in which one wants to encode a
proposition in a term or a proof in a type. We call this situation: the
reverse PAT interpretation.

\begin{itemize}

\item
\textbf{The Automath experience.}\\
Historically the embedding of logic inside $\lambda$-calculus does not
always follow the PAT interpretation. This is the case of $\AUT$ \cite{SPAb1}:
a language of the Automath family \cite{SPAa2} that is very close to a
$\lambda$-calculus. This language has only one kind, named \texttt{type},
and this forces the embedding of logic clearly shown in \cite{SPAd1}, which
is used throughout the formal specification of Landau's Grundlagen
\cite{SPA3}.

We summarize the situation in \figref{embedding}{}.
In $\AUT$ the proofs of a proposition do not inhabit the proposition
directly, as in the PAT interpretation, but they inhabit the 
``assertion type'' associated to the proposition. In this way a proposition
differs from the type of its proofs.

\item
\textbf{The realizability tradition.}\\
One of the basic ideas behind type theory is that terms encode some
entities (for instance computable functions, computer programs, propositions,
proofs) and these entities satisfy a desired property if the corresponding
terms are typable.
In this respect there are type systems set up to capture some properties of
propositions. For instance in the computer program verification scenario one
can state that a proposition is admissible if it the specification of a program
(this idea is taken from the realizability tradition \cite{Kle45}, where the
admissible formulae are those having a realizer, i.e. an implementation). 
In this perspective one may want to encode the propositions in the terms and
their realizers or implementations in the types.
This is the case of $\PML$ \cite{Raf07a,Raf07b,Raf08a}:
an experimental programming language with program verification support.
Notice that in $\PML$ the standard PAT interpretation is also allowed.

\end{itemize} 

\begin{figure}
\begin{center}

\begin{tabular}{|c|c|c|c|}
\hline
\textbf{Encoding}&\textbf{PAT}&\textbf{$\AUT$}&\textbf{$\PML$}\\
\hline
kinds&sort of propositions&\texttt{type}&\\
\hline
types&propositions&sort of propositions, assertion types&realizers\\
\hline
terms&proofs&propositions, proofs&specifications\\
\hline
\end{tabular}

\end{center}
\caption{Different embeddings of logic in type theory}
\figlabel{embedding}
\end{figure}

The above considerations lead to think that a type theory intended as a
logical framework is more flexible if it supports both PAT interpretations
at the same time instead of supporting just one of them (either the standard
one or the inverse one).

This result is achieved by designing the type theory in such a way that both
terms and types are capable of encoding either a proof or a proposition.

The simplest way to obtain this feature is by allowing on one hand the term
constructions at the level of types and on the other hand the type
constructions at the level of terms.
By so doing, we are naturally led to unify terms and types.

It is worth remarking that this unification already appears to some extent in
a number of works including \cite{SPAa2,SPAc3,SPAc4,SPAc6,Cqn85,Kam05}. 

Coming now to the treatment of environments, there are well established
motivations for allowing these structures to contain not just declarations,
but abbreviations (i.e. non-recursive definitions) as well. 
We mention the following ones.

\begin{itemize}

\item
\textbf{Practically unavoidable.}\\
Abbreviations allow to factorize large terms increasing their readability.
It is a matter of fact that Mathematics is unimaginable without abbreviations
and for this reason every type theory designed as a realistic foundation for
developing Mathematics includes some kind of abbreviation mechanism.
Taking three very different examples of such theories, we can mention the
Automath languages \cite{SPAa2}, Constructive Type Theory \cite{NPS90} and 
the Calculus of Inductive Constructions \cite{CP89}.

\item
\textbf{Efficient reduction.}\\
Abbreviations allow to write the $\beta$-contraction in the call-by-name style
\cite{CH00} ``$\MAppl{(\ABST{x}{W}{t})}{v} \SR{\beta} \MAbbr{x}{v}{t}$''
with the effect of delaying the substitution of $v$ in $t$. 
This feature is a crucial ingredient of optimal reduction strategies
\cite{AG99} and is exploited in real reduction machines.

\end{itemize}

Very convenient extensions of well established calculi by means of
abbreviations are presented in \cite{KBN99,CH00}.
 
Once environments are equipped with abbreviations, we see motivations for
pursuing a full duality between environments and terms.

\begin{itemize}

\item 
\textbf{Aggregates without inductive types.}
Aggregate data structures, or aggregates for short, play a central role both
in programming languages (where they appear as records, modules or objects)
and in Mathematics (where they appear as mathematical structures).
The type theories featuring aggregates as terms usually exploit inductive
types for this purpose, but the machinery for supporting inductive types is
too complex if one is only interested in supporting aggregates \cite{Bru91},
especially if dependent types are allowed.
On the other hand every type theory has some support for environments
and an environment with abbreviations can serve as an aggregate with
dependent components.
In this respect we conjecture that supporting environments as terms is much
simpler than supporting inductive types for the only purpose of having
aggregates as terms.

\item 
\textbf{The $\LM$ tradition.}
Beside terms, types and environments, the $\lambda$-calculi for the PAT
interpretation of classical logic derived from $\LM$ \cite{SU06} include
structures called ``contexts'' that play the role of continuations in
functional programming.
The most general of these calculi, $\LBMMT$ \cite{CH00}, features
abbreviations in contexts (but not in terms) and a duality between terms and
contexts, which yet does not yield the unification of the two.
On the other hand we conjecture that contexts can be easily injected into
environments with abbreviations if these environments are also equipped with
other constructions usually found in terms (for instance applications).
Such extended environments become very close to terms themselves and may be
realized by pursuing a ``terms as environments'' discipline in the design of
the type theory. 

\end{itemize}

%% file: outline.tex
\subsection{Outline}
\subseclabel{outline}

This paper describes a typed $\lambda$-calculus, that we call $\LD$ after the
names of its binders, that aims at the unification of terms, types, kinds and
environments both in a static sense and in a dynamic sense.
The static unification lays on the use of a suitable set of expressions that
can represent terms, types, kinds and environments simultaneously.
Additionally, the dynamic unification lays on allowing the same reduction
steps on these expressions whatever they represent.

We are interested in respecting the following desirable constraints:
this calculus must have a well conceived meta-theory, which includes the
commonly required properties and, as a logical framework, must have enough
flexibility and expressive power to encode a non-trivial fragment of
Mathematics in a realistic manner. 

The above considerations imply that the design of $\LD$ involves two crucial
aspects: the choice of the expressions and the choice of the reduction steps
allowed on the expressions. 
In this section we want to discuss these aspects and to analyze their impact
on the capability of our calculus to meet the requirements we have set.

\smallskip

\textbf{The set of the expressions.}
Our approach in this paper is to build expressions using a reasonably small
set of constructions, which we plan to extend in the future.

The starting point is the calculus $\LI$ \cite{SPAc6} where a set $\Delta$ of
expressions is generated by a sort $\tau$, variable occurrences, binary
applications and typed $\lambda$-abstractions in which the types themselves
are expressions in $\Delta$.

This is a very basic platform to which we apply the following modifications.
Firstly we add untyped abbreviations, like ``$\MAbbr{x}{v}{t}$'', following
the motivation outlined in \subsecref{motivations}. 
Secondly we notice that the presence of untyped sorts (as $\tau$ in $\LI$ or
as $\square$ in the $\Cube$ \cite{Brn92}) complicates the meta-theory
unnecessarily because a distinction must be made between the legal expressions
having a type and the legal expressions not having a type. 
To overcome this drawback we use an infinite number of sorts
in place of the single sort $\tau$ and we set up a type system (see below) in
which every sort is typed.
Thirdly we add explicit type annotations (also known as ``explicit type
casts'' in some programming languages) to obtain another meta-theoretical
benefit: with these constructions we easily reduce the type checking problem
to the type inference problem. 

The main limitation of the above set of constructions is the absence of the
higher-order abstraction (i.e. the $\Pi$ construction of the shapes
$\ShapeBS$ and $\ShapeBB$ according to Barendregt's classification), which
essentially sets the expressive power of $\LD$ to that of $\LP$ \cite{Brn92}.%
\footnote{%
Currently we do not have a proof of this statement, but our conjecture is
based on the general consensus that $\LI$ has the expressive power of
$\LP$ \cite{Brn92}.%
}
In any case we can assume that this power is enough to encode non-trivial
parts of Mathematics \cite{SPA3}.%
\footnote{%
We are aware that $\AUTQE$ is a bit more powerful than $\LP$ \cite{KLN04}.%
}

We also set the additional limitation that a variable occurrence is not
an environment constructor because the interpretation of an expression like
``$\ABST{x}{W}{\LREF{x}}$'' as an environment is not straight-forward at all
(here $W$ stands for an expression).
However in \appref{complete} we give some hints on how we plan to face this
problem.

As a consequence we use two sets of expressions, one for the terms (that
also serve as types and kinds) and one for the environments, which is a proper
subset of the former.
This means that $\LD$ realizes the unification of types and terms, which is
the focus of the calculus, but it does not realize the unification of
environments and terms yet.
Namely environments are just expressions formally generated by some term
constructors, but $\LD$ has no support for using them as terms.

It is important to notice that $\LD$ differs from the Automath-related
$\lambda$-calculi \cite{SPA94} in that they do not provide for an abbreviation
construction at the level of terms.
We also notice that when abbreviations are used, the $\lambda$-abstraction it
is not strictly necessary for building a logical framework.
This is the case of $\PAL$ \cite{Luo03}: a platform where partial applications
of functions are not allowed.
As a matter of fact, partial applications have well established benefits in
several contexts including practical functional programming, so our choice is
definitely to include the $\lambda$-abstraction in our calculus.

\smallskip

\textbf{The set of the reduction schemes.}
The reduction schemes aim at realizing deterministic and confluent
computations (as the ones of $\LI$), so critical pairs are avoided for
simplicity.
Since $\LD$ is not focused on achieving the unification of terms and
environments, its reduction schemes work only on terms and no support is given
for the reduction of environment constructors.
Nevertheless these schemes are designed following the principle that they
should also work on environments when possible.
In particular we must be aware that an environment is essentially a list of
declarations (that we represent with $\lambda$-abstractions) and abbreviations
whose position must be preserved when the environment is reduced.

For this reason we use the call-by-name $\beta$-contraction scheme in place
of its call-by-value version (the one used by $\LI$) because the
$\lambda$-abstraction in the redex becomes an abbreviation in the reductum
instead of being deleted.
Another advantage of the call-by-name $\beta$-reduction is discussed in 
\subsecref{motivations}.

Moreover we have three reduction schemes working on abbreviations: namely
a $\delta$-expansion to unfold an abbreviation without removing it,
a $\zeta$-contraction for removing an unreferenced abbreviation (this
reduction would not be allowed if the abbreviation were an environment
constructor) and a $\upsilon$-swap for permuting an application-abbreviation
pair as in \cite{CH00}.

Finally we have a $\tau$-contraction for removing explicit type annotations.

Remarkably we do not consider the $\eta$-contraction.
This is a choice of many calculi including $\LI$ and the systems of the
$\Cube$ \cite{Brn92}.

Also notice that we can obtain a call-by-value $\beta$-contraction by
concatenating a call-by-name $\beta$ contraction, a $\delta$-expansion and a
$\zeta$-contraction.

\smallskip

\textbf{The type system.}
Our aim is to confine the dynamic aspect of the type assignment in the
so-called ``conversion rule'' \cite{Brn92}. This means that we wish to remove
any reference to reduction from the other type assignment rules.
The technical benefit of this approach is that we make clear syntactical
distinction between the construction steps and the conversion steps needed to
infer a type.

\emph{Typed sorts.}
We have a sequence of sorts $h \mapsto \SORT{h}$ (where $h$ ranges over the
set $\NAT$ of the natural numbers) and a function
$\Next{g}{} \oftype \NAT \to \NAT$
that we can choose at will as long as $h < \Next{g}{h}$ holds for every $h$.
In this setting $\SORT{h}$ is typed by $\SORT{\Next{g}{h}}$.

\emph{Typed variable occurrences.}
We exploit the idea that an unreferenced variable needs a legal declaration
only if it is the formal argument of a function, to combine the so-called
``start rule'' and ``weakening rule'' \cite{Brn92} in a simpler rule.

\emph{Typed $\lambda$-abstractions.}
We use the policy of $\LI$, which is known as $\lambda$-typing.
Namely up to conversion, the type of a $\lambda$-abstraction is a
$\lambda$-abstraction.
This policy is adopted by many calculi of the Automath family \cite{SPA94}
and by other calculi including \cite{Kam05,Gro93,Wie99}.

\emph{Typed abbreviations.}
We use the $\lambda$-typing pattern with abbreviations in place of $\lambda$-%
abstractions. This approach yields a uniform typing policy for both binders.

\emph{Typed applications.}
We use the ``compatible'' application rule of \cite{KBN99} with $\lambda$ in
place of $\Pi$, because it does not involve reduction.
By so doing, we strengthen the so-called ``applicability condition''%
\footnote{%
This is the condition that an application must satisfy in order to be legal
or well typed.%
}
with respect to $\LI$, but we conjecture that this is a minor drawback.
For instance the term $t \equiv (x_1\ z)$ is legal in the environment
$\Gamma \equiv 
\typed{x_0}{\ABST{y}{\tau}{y}}, \typed{x_1}{x_0}, \typed{z}{\tau}$
for $\LI$ but not for $\LD$.

\emph{Explicit type annotations.}
We use a ``compatible typing'' policy as well.

\smallskip

\textbf{The meta-theoretical properties.}
One of the aims of the present paper is to show that the design features of
$\LD$ we just described are compatible with the presence of a desirable
meta-theory in the usual sense. The main results are:

\begin{itemize}

\item
the reduction is confluent (Church-Rosser property):
\thcref{prc_props}{pr3_confluence};

\item
the reduction is safe (subject reduction property):
\thcref{ty3_sred}{ty3_sred_pr3};

\item
the typed terms are strongly normalizing:
\thcref{ty3_arity_props}{ty3_sn3}.

\end{itemize}

We also prove other standard properties like the correctness of types, the
uniqueness of types up to reduction and the decidability of type the inference
problem.

We stress that the $\lambda$-abstraction is predicative in that
$\TyT{}{\Gamma}{\ABST{x}{W}{t}}{W}$ never holds.
So $\LD$ can serve as a theory of expressions for the type theories requiring
a meta-language with a predicative abstraction like those in the Marin-L\"of
style \cite{MS05,NPS90,MLS84}

%% file: specification.tex
\subsection{The Certified Specification}
\subseclabel{specification}

The initial version of $\LD$ appears in \cite{Gui06a} where the author outlines
the definitions used in \cite{Gui05} to specify an extension of $\LD$ named
$\CLD$ (see \secref{extension}) in the $\CIC$ (CIC).
Using this encoding it is possible to certify all currently proved properties
of $\CLD$ with the CIC-based proof assistants \textsc{coq} \cite{Coq} and
\textsc{matita} \cite{ASTZ06}.

Following the description of $\LI$ in \cite{SPAc6}, the CIC specification 
exploits position indexes \cite{SPAc2} rather names to represent the
bound variable occurrences. However in this paper we will use names. 

Remarkably $\CLD$ was born and developed in the digital format of
\cite{Gui05}, which is not the formal counterpart of some informal material
previously written on paper (as it happens for most of currently digitalized
Mathematics).
In particular the detailed proofs of the properties of $\CLD$ currently exist
only in their digital version.
Producing a hard copy of these proofs is indeed an interesting challenge
because it requires the implementation of a suitable technology for the
mechanical transformation of digital CIC proof terms into human-readable
proofs written in \LaTeX\ format.%
\footnote{In \cite{PRCICPT} we present an effective procedure for transforming
a CIC proof term is a sequence of basic proof steps. We already implemented
this procedure in the proof assistant \textsc{matita}.}
Our estimation on the length of the hard copy is: 600 pages.

In this paper we outline all proofs of our statements by reporting on the
proof strategy and on the main dependences of each proof. Most proofs are by
induction on the length of a derivation or by cases on the last step of a
derivation. Very often both techniques are applied together. 
This procedure breaks the proof in lot of cases which we do not give the
details of (because they are very easy). 
However we report on the interesting cases giving some hints on how they are
solved.

In \appref{proofs} we give the pointers to the digital proof objects
representing the proofs mentioned in the paper. These proof objects are
available as resources of the \HELM\ (\textsc{helm}) \cite{APSGS03}.

In \appref{progress} we present the main advancements of \cite{Gui05}
at its current state over the description given in \cite{Gui06a}.

%% file: definitions.tex
\section{The Description of $\LD$}
\seclabel{definitions}

In this section we will define $\LD$ in terms of
its grammar (\subsecref{language}),
its reduction rules (\subsecref{reduction-defs}) and
its native type assignment rules (\subsecref{native-defs}).
We will also define some relevant auxiliary notions such as
the static type assignment (\subsecref{static-defs}),
the arity assignment (\subsecref{arity-defs}) and
two preorders on environments (\subsecref{preorders}).
Care was taken to order these topics in a way that takes the reader to the
native type assignment rules as soon as possible.

$\LD$ uses three data types: the set $\NAT$ of the \emph{natural numbers}, 
the set $\TRM$ of the \emph{terms} and the set $\ENV$ of the
\emph{environments}. 
$\NAT$ is used to represent sort indexes (all indexes start at $0$),
$\TRM$ contains the expressions the calculus is about 
(also called pseudo-terms)
and $\ENV$ can be seen as a subclass of $\TRM$.
Although it is not strictly necessary, it is convenient to present $\TRM$ and
$\ENV$ as two distinct data types.

In the presentation of $\LD$ in front of the reader, the term variables are
referenced by name and the names for these variables (i.e. $x$, $y$, $\ldots$)
belong to a data type $\VAR$.

Consistently throughout the presentation, we will be using the following
convention about the names of the meta-variables: $i$, $j$, $h$, $k$ will
range over $\NAT$; $T$, $U$, $V$, $W$ will range over $\TRM$ and $C$, $D$,
$E$, $F$ will range over $\ENV$ or will denote a part of an environment.
We use the Latin capital letters for the term meta-variables following the
untyped $\lambda$-calculus tradition \cite{Brn92}
and we use these letters also for the environment meta-variables, instead of
using the standard Greek capital letters, because we follow the
``environments as terms'' policy pursued by $\LD$.

Lists will also be used (we need them in \subsecref{reduction-props} to prove
the strong normalization theorem). 
The names of variables denoting lists will be overlined: 
like $\overline T$ for a list of terms. 
We will use  $\NIL$ for the empty list and the infix semicolon for
concatenation: like $\CONS{T}{\overline T}$.

In order to avoid the explicit treatment of $\alpha$-conversion, we will
assume that the names of the bound variables and of the free variables are
disjoint in every term, judgement and rule of the calculus (this is known as
the ``Barendregt convention'').

%% file: language.tex
\subsection{The Language}
\subseclabel{language}

Our syntax of terms and environments takes advantage of the so-called
\emph{item notation} \cite{KN96b} because of its well documented benefits. 
When using the item notation of $\lambda$-terms, the operands of an
application are presented in reverse order with respect to standard notation,
i.e. the application of $T$ to $V$ is presented like $(T\ V)$ in standard
notation and like $\APPL{V}{T}$ in item notation.
This means that a $\beta$-redex takes the form $\APPL{V}{\ABST{x}{W}{T}}$
rather than $(\ABST{x}{W}{T}\ V)$.
In this situation the argument $V$ and the abstraction $\ABST{x}{W}{}$ are
close to each other rather than having the body $T$ between them, which can
be very long.
In this sense we believe that this notation, which is almost a constant of the
Automath-related works \cite{SPA94}, improves the visual understanding of
$\beta$-redexes by helping the reader to find the argument-abstraction pairs
more easily.

\begin{definition}[terms and environments]\hfil
\objlabel{TE}

The terms of $\LD$ are made of these syntactical items:
$\SORT{h}$ (sort), $\LREF{x}$ (variable occurrence), $\ABST{x}{W}{}$
(abstractor), $\ABBR{x}{V}{}$ (abbreviator), $\APPL{V}{}$ (applicator) and
$\CAST{W}{}$ (type annotator). 
The sets of terms and environments are defined as follows:
\[\TRM \equiv \SORT{\NAT} \| \LREF{\VAR} \| 
              \ABST{\VAR}{\TRM}{\TRM} \| \ABBR{\VAR}{\TRM}{\TRM} \|
              \APPL{\TRM}{\TRM} \| \CAST{\TRM}{\TRM} 
\]
\[\ENV \equiv \SORT{\NAT} \| 
              \ABST{\VAR}{\TRM}{\ENV} \| \ABBR{\VAR}{\TRM}{\ENV} \|
              \APPL{\TRM}{\ENV} \| \CAST{\TRM}{\ENV}
\objlabel{E}\]
\end{definition}

In the above definition 
$\SORT{h}$ is the sort of index $h$,
$\LREF{x}$ is a variable occurrence,
$\ABST{x}{W}{T}$ is the usual $\lambda$-abstraction (simply abstraction
henceforth) of $T$ over the type $W$,
$\ABBR{x}{V}{T}$ is the abbreviation of $V$ in $T$ (i.e. $\MAbbr{x}{V}{T}$),
$\APPL{V}{T}$ is the application of $T$ to $V$ (i.e. $(T\ V)$ in standard
notation) and
$\CAST{W}{T}$ is the type annotation of $T$ with $W$ (i.e. $(T:W)$ in
\textsc{ml} notation).

We remark that type annotations allow to reduce the type checking problem to
the type inference problem: see \thcref{ty3_gen}{ty3_gen_cast} and
\thcref{ty3_props}{ty3_typecheck}.

We can generalize the application to
$\APPL{V_1 \mathbin; \ldots \mathbin; V_i}{T}$ that denotes
$\APPL{V_1}{} \ldots \APPL{V_i}{T}$. 

It follows from \defcref{TE}{E} that an environment $E$ is always of the form
$\PUSH{C}{\SORT{h}}$, so we allow the notations 
$\PUSH{E}{\ABST{x}{W}{}}$ and $\PUSH{E}{\ABBR{x}{V}{}}$
by which we mean the environments
$\PUSH{C}{\ABST{x}{W}{\SORT{h}}}$ and 
$\PUSH{C}{\ABBR{x}{V}{\SORT{h}}}$ respectively.

A focalized term is an ordered pair $(E, T)$ representing a term $T$ closed
in an environment $E$.
In the ``environments as terms'' perspective pursued by $\LD$, we can also
think that such a pair denotes the concatenation of $T$ after $E$.
Namely $(\PUSH{C}{\SORT{h}}, T)$ may denote the term $\PUSH{C}{T}$.
We stress that focalized terms play an essential role in the substitution
lemma for typing, \thcref{ty3_props}{ty3_fsubst0}, and in the proof that the
type inference problem is decidable, \thcref{ty3_dec}{ty3_inference}.

%% file: helpers.tex
\subsection{Some Helper Operators}

Now we can introduce some operators that we will use in the next sections.

\begin{definition}[free variables]\hfil

The subset $\FV{T}$ contains the free variables occurring in the term $T$. 

The free variables of a term are defined as usual.
\end{definition}

\begin{definition}[environment predicate]\hfil
\objlabel{env}

The predicate $\Env{T}$ states that the term $T$ has the shape of an 
environment.

\begin{itemize}

\item (sort)
$\Env{\SORT{h}}$;

\item (compatibility)
if $\Env{T}$ then \\
$\Env{\ABST{x}{W}{T}}$ and $\Env{\ABBR{x}{V}{T}}$ and
$\Env{\APPL{V}{T}}$ and $\Env{\CAST{W}{T}}$.

\end{itemize}

\end{definition}

We need this predicate only because in $\LD$ some terms are not environments
(see \subsecref{outline}) and we use it just in \thcref{sty_props}{sty1_env}.

The substitution operators we define below are exploited by the current
reduction rules (see \subsecref{reduction-defs}), but we conjecture that
these rules can be reformulated without mentioning substitution explicitly.

\begin{definition}[strict substitution on terms]\hfil
\objlabel{subst0}

The non-deterministic partial function $\SubstZ{y}{W}{T}{}$ substitutes $W$
for one or more occurrences of $y$ in $T$ while it remains undefined if
$y \in \FV{W}$ or if $y \notin \FV{T}$.

The subscript ``t'' is part of the notation and the `+'' recalls
``one or more''.

\begin{enumerate}

\item (var)
if $y\notin\FV{W}$ then $\SubstZ{y}{W}{\LREF{y}}{W}$;

\item (compatibility)
if $\SubstZ{y}{W}{V_1}{V_2}$ and $\SubstZ{y}{W}{T_1}{T_2}$ then

\begin{enumerate}

\item (abst)
$\SubstZ{y}{W}{\ABST{x}{V_1}{T}}{\ABST{x}{V_2}{T}}$ and
$\SubstZ{y}{W}{\ABST{x}{V}{T_1}}{\ABST{x}{V}{T_2}}$ and\\
$\SubstZ{y}{W}{\ABST{x}{V_1}{T_1}}{\ABST{x}{V_2}{T_2}}$;

\item (abbr)
$\SubstZ{y}{W}{\ABBR{x}{V_1}{T}}{\ABBR{x}{V_2}{T}}$ and
$\SubstZ{y}{W}{\ABBR{x}{V}{T_1}}{\ABBR{x}{V}{T_2}}$ \\ and
$\SubstZ{y}{W}{\ABBR{x}{V_1}{T_1}}{\ABBR{x}{V_2}{T_2}}$;

\item (appl)
$\SubstZ{y}{W}{\APPL{V_1}{T}}{\APPL{V_2}{T}}$ and
$\SubstZ{y}{W}{\APPL{V}{T_1}}{\APPL{V}{T_2}}$ and \\
$\SubstZ{y}{W}{\APPL{V_1}{T_1}}{\APPL{V_2}{T_2}}$;

\item (cast)
$\SubstZ{y}{W}{\CAST{V_1}{T}}{\CAST{V_2}{T}}$ and
$\SubstZ{y}{W}{\CAST{V}{T_1}}{\CAST{V}{T_2}}$ and \\
$\SubstZ{y}{W}{\CAST{V_1}{T_1}}{\CAST{V_2}{T_2}}$.

\end{enumerate}

\end{enumerate}

\end{definition}

As already pointed out in \cite{Gui06a}, the function that substitutes $W$ for
$y$ in $T$ can be defined in many different ways. The difference lays in
the number of occurrences of $y$ that a single application of the function
can substitute.
The choices are: one, one or more, zero or more, all, all if one exists.
Our approach is to adopt the second choice and we can justify it with some
technical reasons connected to reduction (see \subsecref{reduction-defs}).
$\LD$ currently defines two $\delta$-reduction rules (i.e. expansions of
local definitions) and we want to use the same substitution function in the
description of both rules. 
This consideration rules out the first choice of the above list because it
invalidates \thcref{prc_props}{pr0_subst0}, that is a prerequisite of
\thcref{prc_props}{pr3_confluence}.
The third and the forth choices, that are the most used in the literature, do
not have this problem, but complicate one of the $\delta$-reduction rules if
we want to preserve its ``orthogonality'' (i.e. absence of critical pairs)
with respect to the $\zeta$-reduction rule. 
Is important to stress that this ``orthogonality'' simplifies the proof of
\thcref{prc_props}{pr0_confluence}: another prerequisite of
\thcref{prc_props}{pr3_confluence}.     
The last choice of the above list is simply too complex with respect to the
benefits it gives. 

Notice that with our substitution function we can not replace a variable
with itself but this is not a problem since we use this function just to
evaluate the $\delta$-redexes, i.e. we use it just to expand non-recursive
definitions.  

Using the same approach, we can define the strict substitution on
environments.

\begin{definition}[strict substitution on environments]\hfil
\objlabel{csubst0}

The non-deterministic partial function $\CsubstZ{y}{W}{E}{}$ substitutes the
term $W$ in the environment $E$ for one or more occurrences of the variable
$y$ occurring in $E$.

The subscript ``$e$'' is part of the notation and the `+'' recalls
``one or more''.

The rules are the following:
if $\SubstZ{y}{W}{V_1}{V_2}$ and $\CsubstZ{y}{W}{E_1}{E_2}$ then

\begin{enumerate}

\item (abst)
$\CsubstZ{y}{W}{\ABST{x}{V_1}{E}}{\ABST{x}{V_2}{E}}$ and
$\CsubstZ{y}{W}{\ABST{x}{V}{E_1}}{\ABST{x}{V}{E_2}}$ and\\
$\CsubstZ{y}{W}{\ABST{x}{V_1}{E_1}}{\ABST{x}{V_2}{E_2}}$;

\item (abbr)
$\CsubstZ{y}{W}{\ABBR{x}{V_1}{E}}{\ABBR{x}{V_2}{E}}$ and
$\CsubstZ{y}{W}{\ABBR{x}{V}{E_1}}{\ABBR{x}{V}{E_2}}$ \\ and
$\CsubstZ{y}{W}{\ABBR{x}{V_1}{E_1}}{\ABBR{x}{V_2}{E_2}}$;

\item (appl)
$\CsubstZ{y}{W}{\APPL{V_1}{E}}{\APPL{V_2}{E}}$ and
$\CsubstZ{y}{W}{\APPL{V}{E_1}}{\APPL{V}{E_2}}$ and \\
$\CsubstZ{y}{W}{\APPL{V_1}{E_1}}{\APPL{V_2}{E_2}}$;

\item (cast)
$\CsubstZ{y}{W}{\CAST{V_1}{E}}{\CAST{V_2}{E}}$ and
$\CsubstZ{y}{W}{\CAST{V}{E_1}}{\CAST{V}{E_2}}$ and \\
$\CsubstZ{y}{W}{\CAST{V_1}{E_1}}{\CAST{V_2}{E_2}}$.

\end{enumerate}

\end{definition}

The strict substitution on focalized terms is defined following the same
pattern.

\begin{definition}[strict substitution on focalized terms]\hfil
\objlabel{fsubst0}

The non-deterministic partial function $\FsubstZ{y}{W}{E}{T}{}{}$ substitutes
$W$ in $(E,T)$ for one or more occurrences of the variable $y$ occurring in
$(E,T)$.

The subscript ``$f$'' is part of the notation and the `+'' recalls
``one or more''.

The rules are the following:
if $\CsubstZ{y}{W}{E_1}{E_2}$ and $\SubstZ{y}{W}{T_1}{T_2}$ then\\
$\FsubstZ{y}{W}{E_1}{T}{E_2}{T}$ and
$\FsubstZ{y}{W}{E}{T_1}{E}{T_2}$ and\\
$\FsubstZ{y}{W}{E_1}{T_1}{E_2}{T_2}$.

\end{definition}

The strict substitution on focalized terms is needed to state the substitution
lemma for the native type assignment in a way that breaks the mutual
dependences existing between the analogous lemmas stated just for the strict
substitution on terms and on environments (see \thcref{ty3_props}{}).

%% file: reduction_defs.tex
\subsection{Reduction and Conversion}
\subseclabel{reduction-defs}

The equivalence of terms in $\LD$ is based on
\emph{environment-dependent conversion}, that is the reflexive, symmetric and
transitive closure of \emph{environment-dependent reduction}.
The latter is expressed in terms of \emph{environment-free reduction}, that is
the compatible closure of five reduction schemes named: $\beta$, $\delta$,
$\zeta$, $\tau$, $\upsilon$.

The purpose of the present section is to describe this construction in detail. 

The need for environment-dependent reduction and conversion derives from the
presence of abbreviations in environments \cite{KBN99}: for example in the
environment $\PUSH{E}{\ABBR{x}{V}{}}$ we want to $\delta$-expand the term $x$
to $V$.

\begin{definition}[environment-free reduction on terms]\hfil
\objlabel{pr0}

The relation $\PrZ{T_1}{T_2}$ indicates one step of environment-free parallel
reduction from $T_1$ to $T_2$. Its rules are in \figref{pr0}{}.
The reduction steps are in \figref{environment-free}{}.
\end{definition}

\begin{figure}
\begin{center}

\Rule{}{\mathrm{refl}}{\PrZ{T}{T}} \quad
\Rule{\PrZ{W_1}{W_2} \quad \PrZ{T_1}{T_2}}{\mathrm{abst}}
     {\PrZ{\ABST{x}{W_1}{T_1}}{\ABST{x}{W_2}{T_2}}} \quad
\Rule{\PrZ{V_1}{V_2} \quad \PrZ{T_1}{T_2}}{\mathrm{abbr}}
     {\PrZ{\ABBR{x}{V_1}{T_1}}{\ABBR{x}{V_2}{T_2}}} \\
\Rule{\PrZ{V_1}{V_2} \quad \PrZ{T_1}{T_2}}{\mathrm{appl}}
     {\PrZ{\APPL{V_1}{T_1}}{\APPL{V_2}{T_2}}} \quad
\Rule{\PrZ{W_1}{W_2} \quad \PrZ{T_1}{T_2}}{\mathrm{cast}}
     {\PrZ{\CAST{W_1}{T_1}}{\CAST{W_2}{T_2}}} \\
\Rule
 {\PrZ{V_1}{V_2} \quad \PrZ{T_1}{T_2}}{\beta}
 {\PrZ{\APPL{V_1}{\ABST{x}{W}{T_1}}}{\ABBR{x}{V_2}{T_2}}} \quad
\Rule
 {\PrZ{V_1}{V_2} \quad \PrZ{T_1}{T_2} \quad \SubstZ{x}{V_2}{T_2}{T}}
 {\delta} {\PrZ{\ABBR{x}{V_1}{T_1}}{\ABBR{x}{V_2}{T}}} \\
\Rule
 {\PrZ{T_1}{T_2} \quad x \notin \FV{T_1}}{\zeta}
 {\PrZ{\ABBR{x}{V}{T_1}}{T_2}} \quad
\Rule{\PrZ{T_1}{T_2}}{\tau}{\PrZ{\CAST{W}{T_1}}{T_2}} \quad
\Rule
 {\PrZ{V_1}{V_3} \quad \PrZ{V_2}{V_4} \quad \PrZ{T_1}{T_2}}{\upsilon}
 {\PrZ{\APPL{V_1}{\ABBR{x}{V_2}{T_1}}}{\ABBR{x}{V_4}{\APPL{V_3}{T_2}}}}

\end{center}
\caption{Environment-free parallel reduction rules on terms}
\figlabel{pr0}
\end{figure}

\begin{figure}
\begin{center}

\begin{tabular}{|l|llll|}
\hline
\textbf{scheme} & \textbf{redex} & & \textbf{reductum} & \\ \hline

$\beta$-contraction & 
$\APPL{V}{\ABST{x}{W}{T}} $ & $\SR{\beta}$ &
$\ABBR{x}{V}{T}$ & \\ \hline

$\delta$-expansion &
$\ABBR{x}{V}{T}$ & $\SR{\delta}$ &
$\ABBR{x}{V}{\SubstZ{x}{V}{T}{}}$ & if $x \in \FV{T}$ \\ \hline

$\zeta$-contraction &
$\ABBR{x}{V}{T}$ & $\SR{\zeta}$ & $T$ & if $x \notin \FV{T}$ \\ \hline

$\tau$-contraction &
$\CAST{W}{T}$ & $\SR{\tau}$ & $T$ & \\ \hline

$\upsilon$-swap &
$\APPL{V_1}{\ABBR{x}{V_2}{T}}$ & $\SR{\upsilon}$ &
$\ABBR{x}{V_2}{\APPL{V_1}{T}}$ & \\ \hline
\end{tabular}

\end{center}
\caption{Environment-free reduction steps}
\figlabel{environment-free}
\end{figure}

Environment-free reduction is presented in its parallel form to ease the proof of
the Church-Rosser property stated by \thcref{prc_props}{pr0_confluence}.
In fact using parallel reduction, we bypass the necessity to trace redexes as
done in \cite{Brn92}.

The effect of a step $\PrZ{T_1}{T_2}$ is to reduce a subset of the redexes
appearing in $T_1$.

The $\beta$ scheme does not perform a full $\beta$-contraction in the usual
sense, but converts a $\beta$-redex into a $\delta$-redex or a $\zeta$-redex,
leaving the rest of the contraction to these two schemes.
The $\delta$ scheme expands (i.e. unfolds) some instances of an abbreviation
(but not necessarily all of them), so the binder remains in place after the
expansion to allow other instances of the same abbreviation to be unfolded
if necessary. 
The $\zeta$ scheme removes the binder of a fully expanded abbreviation
(this can be related to \textsc{coq} \cite{Coq} but the $\zeta$ scheme of
\textsc{coq} unfolds the abbreviation before removing its binder, which we do
by invoking the $\delta$ scheme).
The $\tau$ scheme makes type annotations eliminable up to reduction.
In this way, we express the fact that these items are not strictly essential
for reduction and typing.
The $\upsilon$ scheme is thought to contract the $\beta$-redex
$\APPL{V_1}{\ABST{x}{W}{}}$ when its two items are separated by an extraneous
abbreviator (i.e. $\ABBR{y}{V_2}{}$). Without the $\upsilon$-swap, the
$\beta$-redex would be created only after removing this abbreviator by
$\zeta$-contraction; this means that the associated abbreviation should
be completely unfolded before the removal. 
With the $\upsilon$-swap, instead, we can obtain the $\beta$-redex without
any unfolding and this is certainly more desirable in realistic use cases.

It is worth remarking how the full $\beta$-contraction is achieved in this
calculus: the full $\beta$-contraction performs three atomic actions on the
term $\APPL{V}{\ABST{x}{W}{T}}$:
it removes the applicator, it removes the binder, it substitutes $V$ for
all occurrences of $x$ in $T$.
In $\LD$ special care is taken for having three different reduction schemes
that take charge of these actions.
The $\beta$ scheme is responsible for removing the applicator (the binder
is changed but it is not removed). The substitution is performed by invoking
the $\delta$ scheme one or more times as long as $x$ occurs in $T$.
When the substitution is completed, the $\zeta$ scheme can be applied and the
binder is removed.

As we see, the five reduction schemes are ``orthogonal'' or ``primary'' in
the sense that a given redex belongs to just one scheme and therefore it
reduces in a unique way.
This means that we never have critical pairs. 
Here we are using ``primary'' as opposed to ``auxiliary'' of
\cite{KB05a,KB05b}.
Other primary or auxiliary reduction schemes might be considered as well.

The above reduction allows to define a weak parallel reduction on
environments, which we use to prove the subject reduction results 
\thcref{ty3_sred}{ty3_sred_wcpr0_pr0} and
\thcref{arity_sred}{arity_sred_wcpr0_pr0}.
This reduction is \emph{weak} in the sense that it involves just the terms
appearing in the environment items and not the environment items themselves.

\begin{definition}[weak reduction on environments]\hfil
\objlabel{wcpr0}

The relation $\WcprZ{E_1}{E_2}$ indicates one step of weak parallel reduction
from the environment $E_1$ to the environment $E_2$.
Its rules are shown in \figref{wcpr0}{}.
\end{definition}

\begin{figure}
\begin{center}
\Rule{}{\mathrm{refl}}{\WcprZ{E}{E}} \quad
\Rule{\PrZ{W_1}{W_2} \quad \WcprZ{E_1}{E_2}}{\mathrm{abst}}
     {\WcprZ{\ABST{x}{W_1}{E_1}}{\ABST{x}{W_2}{E_2}}} \quad
\Rule{\PrZ{V_1}{V_2} \quad \WcprZ{E_1}{E_2}}{\mathrm{abbr}}
     {\WcprZ{\ABBR{x}{V_1}{E_1}}{\ABBR{x}{V_2}{E_2}}} \\
\Rule{\PrZ{V_1}{V_2} \quad \WcprZ{E_1}{E_2}}{\mathrm{appl}}
     {\WcprZ{\APPL{V_1}{E_1}}{\APPL{V_2}{E_2}}} \quad
\Rule{\PrZ{W_1}{W_2} \quad \WcprZ{E_1}{E_2}}{\mathrm{cast}}
     {\WcprZ{\CAST{W_1}{E_1}}{\CAST{W_2}{E_2}}}
\end{center}
\caption{Weak parallel reduction rules on environments}
\figlabel{wcpr0}
\end{figure}

\begin{definition}[Environment-dependent parallel reduction]\hfil
\objlabel{prc}

The relation $\PrS{E}{T_1}{T_2}$ indicates one step of environment-dependent
parallel reduction from $T_1$ to $T_2$. Its rules are shown in \figref{pr2}{}
and the reduction steps are shown in \figref{environment-dependent}{}.
Moreover the relation $\PrT{E}{T_1}{T_2}$ is the transitive closure of
$\PrS{}{}{}$ and the relation $\PcT{E}{T_1}{T_2}$ is the symmetric and
transitive closure of $\PrS{}{}{}$, that we call environment-dependent parallel
conversion.
\end{definition}

\begin{figure}
\begin{center}
\Rule{\PrZ{T_1}{T_2}}{\mathrm{free}}{\PrS{E}{T_1}{T_2}} \quad
\Rule
 {\GETL{E}{C_1}{\ABBR{x}{V}{C_2}} \quad \PrZ{T_1}{T_2} \quad
  \SubstZ{x}{V}{T_2}{T}
 }{\delta}
 {\PrS{E}{T_1}{T}}
\end{center}
\caption{Environment-dependent parallel reduction rules}
\figlabel{pr2}
\end{figure}

\begin{figure}
\begin{center}
\begin{tabular}{|l|lllll|}
\hline
\textbf{scheme} & & \textbf{redex} & & \textbf{reductum} & \\ \hline

$\delta$-expansion &
$\GETL{}{C_1}{\ABBR{x}{V}{C_2}} \mathrel\vdash$ & $T$ & $\SR{\delta}$ & 
$\SubstZ{x}{V}{T}{}$ & if $x \in \FV{T}$ \\ \hline

\end{tabular}
\end{center}
\caption{Environment-dependent reduction steps}
\figlabel{environment-dependent}
\end{figure}

Also environment-dependent reduction is presented in its parallel form to ease
the proof of confluence with itself (\thcref{prc_props}{pr3_confluence}).
The effect of a step $\PrS{E}{T_1}{T_2}$ is to reduce a subset of the
environment-free redexes appearing in $T_1$ and, optionally, to expand one or
more instances of a global abbreviation stored in $E$.

We are aware that the $\delta$ rule of \figref{pr2}{} could be improved by
using environment-dependent reduction in place of environment-free reduction
in the second premise.

Finally we discard the widely used notation with the $=$ sign for the
conversion relation because we feel that $=$ should be reserved for a generic
equivalence relation.
We could use $=_{\beta\delta\zeta\tau\upsilon}$ to indicate that
conversion is equality up to the indicated reduction steps, but this notation
does not make clear whether these steps are actually performed sequentially or
in parallel.

We recall that a term is \emph{normal} or \emph{in normal form} \cite{Brn92} 
when it can not be reduced. 
Here we use the following definition of a normal term.

\begin{definition}[normal terms]\hfil

The predicate $\NfS{E}{T}$, stating that the term $T$ is normal with respect
to context-dependent parallel reduction $\PrS{E}{}{}$, is defined as follows.
\[\begin{tabular}{lll}
$\NfS{E}{T_1}$&
iff&
for each $T_2$, $\PrS{E}{T_1}{T_2}$ implies $T_1 = T_2$.
\end{tabular}\]

\end{definition}

Here we are taking into account the fact that $\PrS{E}{}{}$ is a reflexive
relation.

We can also extend the normal form predicate to a list of terms meaning the
conjunction of the predicate applied to each element of the list.

According to \cite{GTL89,Brn92} a term $T$ is \emph{strongly normalizable} if
there is no infinite sequence of reduction steps starting from $T$.

\begin{definition}[strongly normalizable terms]\hfil
\objlabel{sn3}

The predicate $\SnT{E}{T}$, stating that the term $T$ is strongly normalizable
with respect to context-dependent parallel reduction $\PrS{E}{}{}$,
is inductively defined by one clause that is a higher order rule:
\[\hbox{\hskip-0.6em
If for each $T_2$, $T_1 \neq T_2$ and
$\PrT{E}{T_1}{T_2}$ imply $\SnT{E}{T_2}$, then
$\SnT{E}{T_1}$%
}\eqlabel{sing}\]

\end{definition}

Indeed if $\NotPrT{E}{T_1}{T_2}$ for all $T_2 \neq T_1$, then $T_1$ is normal
and $\SnT{E}{T_1}$ holds a fortiori.
This is the base case of the structural induction defined by \ruleref{sing}.

Essentially we borrowed this definition from \cite{LS04} but we had to take
into account the fact that $\PrT{E}{}{}$ is a reflexive relation.
Moreover we would prefer to use $\PrS{E}{}{}$ in place of $\PrT{E}{}{}$ but
$\PrS{E}{}{}$ is not perfectly designed yet and some desirable properties fail
to hold: for instance even if $\PrS{E}{V_1}{V_2}$ and $\PrS{E}{T_1}{T_2}$, it
is not true that $\PrS{E}{\APPL{V_1}{T_1}}{\APPL{V_2}{T_2}}$.

We can also extend the strong normalization predicate to a list of terms
meaning the conjunction of the predicate applied to each element of the
list.

%% file: native_defs.tex
\subsection{Native Type Assignment}
\subseclabel{native-defs}

In this subsection we present the native type system of $\LD$.
Another type system, originally due to de Bruijn, is presented in
\subsecref{static-defs}.

The type judgement depends on the parameter defined below:

\begin{definition}[sort hierarchy parameter]\hfil
\objlabel{G}

The sort hierarchy parameter is a function $g \oftype \NAT \to \NAT$ that
satisfies the strict monotonicity condition: $h < \Next{g}{h}$ for all $h$.

\end{definition}

The value $\Next{g}{h}$ is the index of the sort that types $\SORT{h}$ and
the monotonicity of $\Next{g}{}$ is the simplest condition ensuring a loop-free
type hierarchy of sorts. 
We use this condition to prove \thcref{ty3_arity_props}{ty3_acyclic}
(impossibility of typing a term with itself).

Notice that $\Next{g}{}$ is a total function but in the most general case a
partial function should be used.
This would allow sort hierarchies with top-level elements as the ones of many
typed $\lambda$-calculi.
Nevertheless this generalization is inconvenient since it complicates several
theorems about typing without increasing the expressiveness of the calculus,
in fact any sort hierarchy with top-level elements can be embedded in a sort
hierarchy without top-level elements.

\begin{definition}[native type assignment]\hfil 
\objlabel{ty3}

The native type judgement has the form $\TyT{g}{E}{T}{U}$ where $g$ is a sort
hierarchy parameter. Its rules are shown in \figref{ty3}{}.

\end{definition}

\begin{figure}
\begin{center}

\Rule{}{\mathrm{sort}}{\TyT{g}{E}{\SORT{h}}{\SORT{\Next{g}{h}}}}

\Rule
 {\GETL{E}{C_1}{\ABBR{x}{V}{C_2}} \quad \TyT{g}{C_1}{V}{W}}{\mathrm{def}}
 {\TyT{g}{E}{\LREF{x}}{W}} \quad
\Rule
 {\GETL{E}{C_1}{\ABST{x}{W}{C_2}} \quad \TyT{g}{C_1}{W}{V}}{\mathrm{decl}}
 {\TyT{g}{E}{\LREF{x}}{W}}

\Rule
 {\TyT{g}{E}{V}{W} \quad \TyT{g}{\PUSH{E}{\ABBR{x}{V}{}}}{T}{U}}
 {\mathrm{abbr}}{\TyT{g}{E}{\ABBR{x}{V}{T}}{\ABBR{x}{V}{U}}} \quad
\Rule
 {\TyT{g}{E}{W}{V} \quad \TyT{g}{\PUSH{E}{\ABST{x}{W}{}}}{T}{U}}
 {\mathrm{abst}}{\TyT{g}{E}{\ABST{x}{W}{T}}{\ABST{x}{W}{U}}}

\Rule
 {\TyT{g}{E}{V}{W} \quad \TyT{g}{E}{T}{\ABST{x}{W}{U}}}{\mathrm{appl}}
 {\TyT{g}{E}{\APPL{V}{T}}{\APPL{V}{\ABST{x}{W}{U}}}} \quad
\Rule
 {\TyT{g}{E}{T}{W} \quad \TyT{g}{E}{W}{V}}{\mathrm{cast}}
 {\TyT{g}{E}{\CAST{W}{T}}{\CAST{V}{W}}}

\Rule
 {\TyT{g}{E}{U_2}{W} \quad \TyT{g}{E}{T}{U_1} \quad \PcT{E}{U_1}{U_2}}
 {\mathrm{conv}}{\TyT{g}{E}{T}{U_2}}

\end{center}
\caption{Native type assignment rules}
\figlabel{ty3}
\end{figure}

Notice that the $\LD$ type judgement does not depend on the notion of a legal
(i.e. well formed) context as it happens in other type systems (see for
instance \cite{MS05}).
This is because an unreferenced variable needs a legal declaration only if it
is the formal argument of a function.
This approach, which is closer to a realistic implementation of a type
checker, has the technical benefit of simplifying the proofs of the properties
of types because the mutual dependence between the type judgement and the
legality judgement disappears.

The type policy of $\LD$ is that the type rules should be as close as
possible to the usual rules of typed $\lambda$-calculus \cite{Brn92}. 
The major modification lays in the type rule for abstraction, that is the
composition of the usual type rules for $\lambda$ and for $\Pi$.
Here are the type rules for $\lambda$ and for $\Pi$ in the $\lambda$-cube.
\[
\Rule
 {\Gamma, x\mathord:A \vdash b \mathrel: B \qquad
  \Gamma \vdash (\Pi_{x:A}.B) \mathrel: s}{}
 {\Gamma \vdash (\lambda_{x:A}.b) \mathrel: (\Pi_{x:A}.B)}
\qquad
\Rule
 {\Gamma \vdash A \mathrel: s_1 \qquad
  \Gamma, x\mathord:A \vdash B \mathrel: s_2}{}
 {\Gamma \vdash (\Pi_{x:A}.B) \mathrel: s_2}
\eqlabel{lambdapi}\]

In $\LD$ we want to type an abstraction with an abstraction, therefore we
remove the second premise of the first rule and the conclusion of the second
rule. Then we make a single rule by combining the remaining judgements and by
turning the $\Pi$ into a $\lambda$.
In addition we generalize the sorts $s_1$ and $s_2$ to arbitrary types.
Moreover we recently noticed that the second premise of the second rule
becomes unnecessary. The rule we obtain at the end is \figref{ty3}{abst}.
An important consequence of this rule, expressed by
\thcref{ty3_arity_props}{ty3_arity},
is that a term and its type have the same functional structure, i.e.
they take the same number of arguments when they are interpreted as functions,
moreover the corresponding arguments of these functions have the same type.
Stated in other words, a type fully determines the number of arguments
taken by its inhabitants and the types of these arguments. 

\figref{ty3}{abbr} follows the scheme of \figref{ty3}{abst} and is
compatible with the commonly accepted \ruleref{delta} for typing abbreviations
found in \cite{Coq} since $B[x:=A]$ and $(\delta_{x=A}.B)$ are
$\delta\zeta$-convertible.
Notice that $C$ does not need to be a sort in this rule.
\[
\Rule
 {\Gamma, x\mathord=A \vdash b \mathrel: B \qquad
  \Gamma \vdash A \mathrel: C}{}
 {\Gamma \vdash (\delta_{x=A}.b) \mathrel: B[x:=A]}
\eqlabel{delta}\]

In the spirit of \figref{ty3}{abbr}, the rule typing the application 
(\figref{ty3}{appl} that we borrow from \cite{KBN99})
does not apply any reduction at the level of types (like \ruleref{delta} does,
unfolding the abbreviation in the term $B$).

The technical benefit of this approach is that the reductional behavior of the
type judgement is confined in the so-called ``conversion rule''.

More sophisticated forms of typing, involving reductions in the environment
(in the sense of \subsecref{reduction-defs}) might be considered as well.

%% file: static_defs.tex
\subsection{Static Type Assignment}
\subseclabel{static-defs}

The so-called \emph{de Bruijn type assignment} ($\mathrm{typ}$ in \cite{Bru93}
and in the Automath tradition) is a function introduced by de Bruijn as part
of the type checking algorithm for the language $\AUT$.
Here we define the analogous concept in $\LD$.

\begin{definition}[Static type assignment]\hfil
\objlabel{sty0}

The partial function $\StyZ{g}{E}{T}{}$ evaluates the static type of a
term $T$ in the environment $E$, which depends on the parameter $g$.
Its rules are shown in \figref{sty0}{}.

The non-deterministic partial function $\StyF{g}{E}{T}{}$ evaluates the
composition of one or more applications of $\StyZ{g}{}{}{}$ to $T$ in $E$.
The `$+$'' recalls ``one or more''.

\end{definition}

\begin{figure}
\begin{center}

\Rule
 {}{\mathrm{sort}}{\StyZ{g}{E}{\SORT{h}}{\SORT{\Next{g}{h}}}} \\
\Rule
 {\GETL{E}{C_1}{\ABBR{x}{V}{C_2}} \quad \StyZ{g}{C_1}{V}{W}}{\mathrm{def}}
 {\StyZ{g}{E}{\LREF{x}}{W}} \quad
\Rule
 {\GETL{E}{C_1}{\ABST{x}{W}{C_2}} \quad \StyZ{g}{C_1}{W}{V}}{\mathrm{decl}}
 {\StyZ{g}{E}{\LREF{x}}{W}} \\
\Rule
 {\StyZ{g}{\PUSH{E}{\ABBR{x}{V}{}}}{T}{U}}{\mathrm{abbr}}
 {\StyZ{g}{E}{\ABBR{x}{V}{T}}{\ABBR{x}{V}{U}}} \quad
\Rule
 {\StyZ{g}{\PUSH{E}{\ABST{x}{W}{}}}{T}{U}}{\mathrm{abst}}
 {\StyZ{g}{E}{\ABST{x}{W}{T}}{\ABST{x}{W}{U}}} \\
\Rule
 {\StyZ{g}{E}{T}{U}}{\mathrm{appl}}
 {\StyZ{g}{E}{\APPL{V}{T}}{\APPL{V}{U}}} \quad
\Rule
 {\StyZ{g}{E}{V}{W} \quad \StyZ{g}{E}{T}{U}}{\mathrm{cast}}
 {\StyZ{g}{E}{\CAST{V}{T}}{\CAST{W}{U}}}

\end{center}
\caption{Static type assignment rules}
\figlabel{sty0}
\end{figure}

Notice that this type is assigned by means of syntax-oriented rules that do
not involve reduction, that is why we term this type \emph{static} in this
paper.

Obviously this feature makes the computation of the static type very fast.
Another consequence is that the static type of a term inherits the binders and
redexes of that term (i.e. it may have more binders and redexes but not less).

Besides being a very well established notion that also $\LD$ can deal with,
the static type is relevant in this paper for two theoretical reasons.
Firstly it allows to define an immersion of $\TRM$ into $\ENV$ that opens the
road to a dualization of terms and environments (see \appref{duality}). 
Secondly it is used in \subsecref{arity-defs} to justify the notion of arity,
that plays an important role in connecting $\LD$ to $\LR$.

%% file: arity_defs.tex
\subsection{Arity Assignment}
\subseclabel{arity-defs}

The notion of arity \cite{NPS90} (skeletons in \cite{Brr96}) as a description
of the functional structure of a term it is not strictly necessary in $\LD$
as well as the data type $\LEV$ used to represent it (since arities can be
encoded into terms). But both are useful from the technical standpoint.
Arities are expected to provide for a connection between the terms of $\LD$
and the types of a suitable version of $\LR$, they facilitate the proof of the
strong normalization theorem (see \thcref{sn3_props}{sc3_arity}) and they
speed up the proofs of the last three clauses of \thcref{ty3_arity_props}{}.

\begin{definition}[arities]\hfil
\objlabel{L}

The set of arities is defined as follows:
\[\LEV \equiv \NODE{\NAT}{\NAT} \| \IMPL{\LEV}{\LEV}\]

The arities of the form $\NODE{k}{h}$ are called nodes and are ordered pairs.

\end{definition}

In the following, the variable $L$ will always range over the data type
$\LEV$.

The arity of a term $T$ has the form
$L \equiv \IMPL{L_1}{\IMPL{L_2}{\IMPL{\ldots}{\IMPL{L_i}{\NODE{k}{h}}}}}$
and it describes the following features of $T$:

\begin{itemize}

\item
the \emph{position} of $T$ in the type hierarchy is the node $\NODE{k}{h}$.
By this we mean that iterating $k$ times the static typing operation on $T$,
we obtain a term whose rightmost item is $\SORT{h}$
(this term exists as shown by \thcref{sty_props}{sty1_env});

\item
$T$ is a function taking exactly $i$ arguments 
(i.e. a function of arity $i$);  

\item
for each $j$ between $1$ and $i$, the $j$-th argument of $T$ must have arity
$L_j$.

\end{itemize}

By looking at its shape, it should be clear that an arity is a type of
the instance of $\LR$ in which we take the nodes as basic types.

Notice that our arity of $T$, containing the position of all arguments of
$T$, is more informative than the skeleton of \cite{Brr96} that only records
the position of $T$.

Also notice that we can not expect a term to have a unique position since
each term at position $\NODE{k}{h}$ is also at position
$\NODE{k+1}{\Next{g}{h}}$.%
\footnote{The converse is not true in general.}

In order to assign an arity to a declared variable we need a function
connecting the arity of a term to the arity of its type. 
Here we present the \emph{strict successor} function defined below
but we are not positive on the fact that this is the best choice and we see
two alternatives that might be considered as well.  

The strict successor of a node depends on the sort hierarchy parameter $g$ and
the strict successor of an arity is a natural extension of the former. 
We also introduce the \emph{strict sum} as the iterated composition of the
strict successor. 

\begin{definition}[the strict successor and the strict sum]\hfil
\objlabel{asucc}

The strict successor of the arity $L$, denoted by $\Asucc{g}{L}$ is defined
as follows:

\[\left\{\begin{tabular}{lll}
$\Asucc{g}{\NODE{0}{h}}$&
$\equiv$&
$\NODE{0}{\Next{g}{h}}$
\\
$\Asucc{g}{\NODE{k + 1}{h}}$&
$\equiv$&
$\NODE{k}{h}$
\\
$\Asucc{g}{(\IMPL{L_1}{L_2})}$&
$\equiv$&
$\IMPL{L_1}{(\Asucc{g}{L_2})}$
\\
\end{tabular}\right.\]
The strict sum $\Aplus{g}{L}{k}$ is the composition of $k$ strict
successors applied to $L$. 
\[\left\{\begin{tabular}{lll}
$\Aplus{g}{L}{0}$&
$\equiv$&
$L$
\\
$\Aplus{g}{L}{(k + 1)}$&
$\equiv$&
$\Asucc{g}{(\Aplus{g}{L}{k})}$
\\
\end{tabular}\right.\]

\end{definition}

We may think of the type hierarchy induced by the parameter $g$ as an
oriented graph in which the arcs are drown from each node $L$ to its
strict successor $\Asucc{g}{L}$.

Coming now to the problem of defining the level (class in \cite{Brr96}) of a
node in the type hierarchy graph, i.e. the height of this node from a
reference point, we observe that this notion can not be given in absolute
terms (as it happens in the type hierarchies with top-level elements or
bottom-level elements) because in our case the graph can be disconnected so no
node can be taken as a global reference point.
The best we can do is to define what it means for two nodes $L_1$ and $L_2$
to be \emph{at the same level} by saying that they must have the same height
relatively to a third node $L_3$ to which they are both connected. 

So we say that the nodes $L_1$ and $L_2$ are at the same level in
the type hierarchy if there exists $k$ such that 
$\Aplus{g}{L_1}{k} = \Aplus{g}{L_2}{k}$
and we express this concept as follows. 

\begin{definition}[level quality]\hfil
\objlabel{leq}

The level equality predicate $\Leq{g}{L_1}{L_2}$ is defined by the rules in
\figref{leq}{}.

\end{definition}

\begin{figure}
\begin{center}

\Rule
 {\Aplus{g}{\NODE{k_1}{h_1}}{k} = \Aplus{g}{\NODE{k_2}{h_2}}{k}}
 {\mathrm{node}}{\Leq{g}{\NODE{k_1}{h_1}}{\NODE{k_2}{h_2}}} \quad
\Rule
 {\Leq{g}{L_1}{L_2} \quad \Leq{g}{L_3}{L_4}}{\mathrm{impl}}
 {\Leq{g}{\IMPL{L_1}{L_3}}{\IMPL{L_2}{L_4}}}

\end{center}
\caption{Level equality rules}
\figlabel{leq}
\end{figure}

Notice that $\Leq{g}{}{}$ is an equivalence relation and that
$\Leq{g}{\NODE{k}{h}}{\NODE{k+1}{\Next{g}{h}}}$ in fact
$\Aplus{g}{\NODE{k}{h}}{(k+1)} = 
 \NODE{0}{\Next{g}{h}} =
 \Aplus{g}{\NODE{k+1}{\Next{g}{h}}}{(k+1)}
$.

Formally the levels of the type hierarchy are the equivalence classes of
$\Leq{g}{}{}$.

If we chose $g(h) \equiv h + 1$, the levels of the corresponding type
hierarchy are isomorphic to the integer numbers, as shown by
\thcref{levels_ex0}{}, and the integer number associated to the equivalence
class containing the node $\NODE{k}{h}$ is $h-k$.
This result is consistent with the intuition according to which the type
hierarchy of $\LD$ has an infinite sequence of levels both above and below
any reference point.%
\footnote{If we define $\Aplus{g}{\NODE{k}{h}}{z} \equiv \NODE{k-z}{h}$
when $z < 0$, then the function $z \mapsto \Aplus{g}{L}{z}$ from the integer
numbers to $\LEV$ is injective with respect to $\Leq{g}{}{}$ in the sense that
$\Leq{g}{\Aplus{g}{L}{z_1}}{\Aplus{g}{L}{z_2}}$ implies $z_1 = z_2$.
This fact is not proved in \cite{Gui05} yet.}

It is important to remark that the decidability of the predicate $\Leq{g}{}{}$
depends on the choice of the parameter $g$. This predicate is undecidable in
general but it is decidable for some choices of $g$, for instance for the one
above. 

Now we have all the ingredients to define the arity assignment.

\begin{definition}[arity assignment]\hfil
\objlabel{arity}

The arity assignment predicate is $\Arity{g}{E}{T}{L}$ and means that the term
$T$ has arity $L$ in the context $E$ with respect to $g$. 
Its rules are given in \figref{arity}{}. 

\end{definition}

\begin{figure}
\begin{center}

\Rule
 {\Arity{g}{E}{T}{L_1} \quad \Leq{g}{L_1}{L_2}}
 {\mathrm{repl}}{\Arity{g}{E}{T}{L_2}} \quad
\Rule
 {}{\mathrm{sort}}{\Arity{g}{E}{\SORT{h}}{\NODE{0}{h}}} \\
\Rule
 {\GETL{E}{C_1}{\ABBR{x}{V}{C_2}} \quad \Arity{g}{C_1}{V}{L}}
 {\mathrm{def}}{\Arity{g}{E}{\LREF{x}}{L}} \quad
\Rule
 {\GETL{E}{C_1}{\ABST{x}{W}{C_2}} \quad \Arity{g}{D}{W}{\Asucc{g}{L}}}
 {\mathrm{decl}}{\Arity{g}{E}{\LREF{x}}{L}} \\
\Rule
 {\Arity{g}{E}{V}{L_1} \quad \Arity{g}{\PUSH{E}{\ABBR{x}{V}{}}}{T}{L_2}}
 {\mathrm{abbr}}{\Arity{g}{E}{\ABBR{x}{V}{T}}{L_2}} \quad
\Rule
 {\Arity{g}{E}{V}{\Asucc{g}{L_1}} \quad
  \Arity{g}{\PUSH{E}{\ABST{x}{W}{}}}{T}{L_2}
 }{\mathrm{abst}}{\Arity{g}{E}{\ABST{x}{W}{T}}{\IMPL{L_1}{L_2}}} \\
\Rule
 {\Arity{g}{E}{V}{L_1} \quad \Arity{g}{E}{T}{\IMPL{L_1}{L_2}}}
 {\mathrm{appl}}{\Arity{g}{E}{\APPL{V}{T}}{L_2}} \quad
\Rule
 {\Arity{g}{E}{W}{\Asucc{g}{L}} \quad \Arity{g}{E}{T}{L}}
 {\mathrm{cast}}{\Arity{g}{E}{\CAST{W}{T}}{L}}

\end{center}
\caption{Arity assignment rules}
\figlabel{arity}
\end{figure}

In this paper we assign the arity up to level equality, but we suspect that
other (more desirable) solutions are possible as well.

%% file: preorders.tex
\subsection{Domain-Based Preorders on Environments}
\subseclabel{preorders}

We recall that a variable occurrence $x$ is a placeholder for a member of a 
given subset of terms, which is called the \emph{domain} of $x$.
In our case if $x$ is bound in the environment
$E_1 \equiv \PUSH{C}{\ABST{x}{W}{}}$
then $x$ stands for any term of type $W$ in $C$ so its domain is
$\DOM_1 \equiv \css{T \st \TyT{g}{C}{T}{W}}$.
On the other hand if $x$ is bound in the environment
$E_2 \equiv \PUSH{C}{\ABBR{x}{V}{}}$
then $x$ stands only for $V$ so its domain is
$\DOM_2 \equiv \css{T \st T=V}$.

If we now assume $\TyT{g}{C}{V}{W}$, we see that $\DOM_2 \subseteq \DOM_1$
and we are led to define the following preorder $\Csubt{g}{}{}$ on
environments such that $\Csubt{g}{E_1}{E_2}$ holds.

\begin{definition}[domain-based preorder on environments]\hfil
\objlabel{csubt}

The relation $\Csubt{g}{E_1}{E_2}$ holds when the environments $E_2$ and $E_1$
bind the same variables and for each of these variables, its domain in $E_2$
is contained in its domain in $E_1$.%
\footnote{In \cite{Gui05} we axiomatized the relation 
``$E_1 \succeq_g E_2$'' rather than ``$\Csubt{g}{E_1}{E_2}$''.}
The rules of this relation are given below: 

\begin{itemize}

\item (sort)
$\Csubt{g}{\SORT{h}}{\SORT{h}}$;

\item (compatibility)
if $\Csubt{g}{C_1}{C_2}$ then
$\Csubt{g}{\ABST{x}{W}{C_1}}{\ABST{x}{W}{C_2}}$ and\\
$\Csubt{g}{\ABBR{x}{V}{C_1}}{\ABBR{x}{V}{C_2}}$ and
$\Csubt{g}{\APPL{V}{C_1}}{\APPL{V}{C_2}}$ and
$\Csubt{g}{\CAST{W}{C_1}}{\CAST{W}{C_2}}$;

\item (abst)
if $\Csubt{g}{C_1}{C_2}$ and $\TyT{g}{C_2}{V}{W}$ and $\TyT{g}{C_1}{V}{W}$
then\\$\Csubt{g}{\PUSH{C_1}{\ABST{x}{W}{}}}{\PUSH{C_2}{\ABBR{x}{V}{}}}$.

\end{itemize}

\end{definition}

The preorder $\Csubt{g}{}{}$ is an auxiliary notion we use to prove the
subject reduction property of the native type assignment,
\thcref{ty3_sred}{ty3_sred_wcpr0_pr0}, in the case of the $\beta$-contraction
because of the shapes of the $\beta$-reductum (\figref{environment-free}{}),
of \figref{ty3}{abst} and of \figref{ty3}{abbr}.
In fact we know that the calculi in which the $\beta$-reductum exploits an
explicit substitution in place of an abbreviation, do not need this apparatus.

If we relax the minor premises of \defref{csubt}{abst} by expressing them
in terms of the arity assignment, we obtain the preorder $\Csuba{g}{}{}$
defined below:

\begin{definition}[relaxed domain-based preorder on environments]\hfil
\objlabel{csuba}

The relation $\Csuba{g}{E_1}{E_2}$ is defined like $\Csubt{g}{E_1}{E_2}$ but
\defref{csubt}{abst} is replaced by the following axiom:

\begin{itemize}

\item (abst)
if $\Csuba{g}{C_1}{C_2}$ and
$\Arity{g}{C_2}{V}{L}$ and $\Arity{g}{C_1}{W}{\Asucc{g}{L}}$ then\\
$\Csuba{g}{\PUSH{C_1}{\ABST{x}{W}{}}}{\PUSH{C_2}{\ABBR{x}{V}{}}}$.

\end{itemize}

\end{definition}

We use this preorder as an auxiliary notion to prove the subject reduction
property of the arity assignment, \thcref{arity_sred}{arity_sred_wcpr0_pr0},
in the case of the $\beta$-contraction because of the shapes of the
$\beta$-reductum (\figref{environment-free}{}), of \figref{arity}{abst} and of
\figref{arity}{abbr}.
We stress that $\Csuba{g}{}{}$ is undecidable in general because it involves
$\Leq{g}{}{}$. 

Notice that \thcref{ty3_arity_props}{csubt_csuba} states that
$\Csubt{g}{E_1}{E_2}$ implies $\Csuba{g}{E_1}{E_2}$ but we argue from 
\thcref{ty3_ex1}{} that the converse does not hold in general.

%% file: theory.tex
\section{The Theory of $\LD$}
\seclabel{theory}

In this section we present the main properties of the notions we introduced
in \secref{definitions}.
In particular we give the results on arities (\subsecref{arity-props}),
on reduction and conversion (\subsecref{reduction-props}),
on native types (\subsecref{native-props}) and
on static types (\subsecref{static-props}).
Notice that here we are forced to order the topics in a slightly different way 
with respect to \secref{definitions} because we want to follow the dependency
graph of the theorems we present.
In \subsecref{examples} we give some theorems about concrete terms and
instances of the parameter $g$ having interesting properties.

%% file: arity_props.tex
\subsection{Results on the Arity Assignment}
\subseclabel{arity-props}

The arity assignment is an auxiliary notion in $\LD$, that we mainly
introduced just to reduce the strong normalization of $\LD$ to that of $\LR$.
Furthermore the \emph{replacement} arity assignment rule,
\figref{arity}{repl}, its not satisfactory because it involves the level
equality predicate, which is undecidable in general. 
For these reasons we prefer not to insist on the results on arities and we
just give some examples below.

\begin{theorem}[main properties of arities]\hfil
\objlabel{arity_props}

\begin{enumerate}\EnumStyle

\item (every node is inhabited)
\objlabel{node_inh}

For all $h$, $k$
there exist $C$, $T$ such that
$\Arity{g}{C}{T}{\NODE{k}{h}}$.

\item (uniqueness of arity up to level equality)
\objlabel{arity_mono}

If $\Arity{g}{C}{T}{L_1}$ and
$\Arity{g}{C}{T}{L_2}$ then
$\Leq{g}{L_1}{L_2}$.

\item (substitution in focalized terms preserves the arity)
\objlabel{arity_fsubst0}

If $\Arity{g}{C_1}{T_1}{L}$ and
$\GETL{C_1}{E}{\ABBR{x}{V}{E\p}}$ and
$\FsubstZ{x}{V}{C_1}{T_1}{C_2}{T_2}$ then
$\Arity{g}{C_2}{T_2}{L}$.

\item (monotonicity of the arity assignment with respect to $\Csuba{g}{}{}$)
\objlabel{csuba_arity}

If $\Arity{g}{C_1}{T}{L}$ and
$\Csuba{g}{C_1}{C_2}$ then
$\Arity{g}{C_2}{T}{L}$.

\end{enumerate}

\end{theorem}

\begin{proof}
Clause \thcref{}{node_inh} is proved by induction on $k$.
Clause \thcref{}{arity_mono} is proved by induction on the first premise and by
cases on the second premise.
Clause \thcref{}{arity_fsubst0} is proved by induction on the first premise and
by cases on the third premise.
Clause \thcref{}{csuba_arity} is proved by induction on the first premise with
some invocations of Clause \thcref{}{arity_mono}.
\end{proof}


The subject reduction property of the arity assignment is proved by the
theorem below.
The main part of the proof is in the base case, where a single step of
environment-free parallel reduction is considered.
The possibility to reduce some terms inside the environment is essential here.
The general case is just a corollary.
As a consequence, the level of a term in the type hierarchy is preserved by
reduction.

\begin{theorem}[subject reduction]\hfil
\objlabel{arity_sred}

\begin{enumerate}\EnumStyle

\item (base case)
\objlabel{arity_sred_wcpr0_pr0}

If $\Arity{g}{C_1}{T_1}{L}$ and
$\WcprZ{C_1}{C_2}$ and
$\PrZ{T_1}{T_2}$ then
$\Arity{g}{C_2}{T_2}{L}$.

\item (general case without the reduction in the environment)
\objlabel{arity_sred_pr3}

If $\PrT{C}{T_1}{T_2}$ and
$\Arity{g}{C}{T_1}{L}$ then
$\Arity{g}{C}{T_2}{L}$.

\end{enumerate}

\end{theorem}

\begin{proof}
Clause \thcref{}{arity_sred_wcpr0_pr0} is proved by double induction on the
first and third premise.
In the case of \figref{arity}{abbr} against \figref{pr0}{\delta} we exploit 
\thcref{arity_props}{arity_fsubst0}, and in the case of \figref{arity}{appl}
against \figref{pr0}{\beta} we exploit \thcref{arity_props}{csuba_arity}.
Clause \thcref{}{arity_sred_pr3} is proved by induction on the first premise via
the previous clause.
\end{proof}

%% file: reduction_props.tex
\subsection{The Results on Reduction and Conversion}
\subseclabel{reduction-props}

The most relevant properties of reduction and conversion are listed below.

\begin{theorem}[main properties of reduction and conversion]\hfil
\objlabel{prc_props} 

\begin{enumerate}\EnumStyle

\item (confluence of $\PrZ{}{}{}$ with strict substitution)
\objlabel{pr0_subst0}

If $\PrZ{T_1}{T_2}$ and
$\SubstZ{x}{W_1}{T_1}{U_1}$ and
$\PrZ{W_1}{W_2}$ then
$\PrZ{U_1}{T_2}$ or
there exists $U_2$ such that
$\PrZ{U_1}{U_2}$ and
$\SubstZ{x}{W_2}{T_2}{U_2}$.

\item (confluence of $\PrZ{}{}{}$ with itself: Church-Rosser property)
\objlabel{pr0_confluence}

If $\PrZ{T_0}{T_1}$ and
$\PrZ{T_0}{T_2}$ then
there exists $T$ such that
$\PrZ{T_1}{T}$ and
$\PrZ{T_2}{T}$.

\item (confluence of $\PrT{}{}{}$ with itself: Church-Rosser property)
\objlabel{pr3_confluence}

If $\PrT{C}{T_0}{T_1}$ and
$\PrT{C}{T_0}{T_2}$ then
there exists $T$ such that
$\PrT{C}{T_1}{T}$ and
$\PrT{C}{T_2}{T}$.

\item (thinning of the applicator for $\PcT{}{}{}$)
\objlabel{pc3_thin_dx_appl}

If $\PcT{C}{T_1}{T_2}$ then
$\PcT{C}{\APPL{V}{T_1}}{\APPL{V}{T_2}}$.

\item (compatibility for $\PcT{}{}{}$: first operand)
\objlabel{pc3_head_1}

If $\PcT{C}{V_1}{V_2}$ then
$\PcT{C}{\ABST{x}{V_1}{T}}{\ABST{x}{V_2}{T}}$ and
$\PcT{C}{\APPL{V_1}{T}}{\APPL{V_2}{T}}$ and
$\PcT{C}{\ABBR{x}{V_1}{T}}{\ABBR{x}{V_2}{T}}$ and
$\PcT{C}{\CAST{V_1}{T}}{\CAST{V_2}{T}}$.

\item (compatibility for $\PcT{}{}{}$: second operand)
\objlabel{pc3_head_2}

If $\PcT{\PUSH{C}{\ABST{x}{V}{}}}{T_1}{T_2}$ then
$\PcT{C}{\ABST{x}{V}{T_1}}{\ABST{x}{V}{T_2}}$;
if $\PcT{\PUSH{C}{\ABBR{x}{V}{}}}{T_1}{T_2}$ then
$\PcT{C}{\ABBR{x}{V}{T_1}}{\ABBR{x}{V}{T_2}}$.

\item (generation lemma on abstraction for $\PcT{}{}{}$)
\objlabel{pc3_gen_abst}

If $\PcT{C}{\ABST{x}{V_1}{T_1}}{\ABST{x}{V_2}{T_2}}$ then
$\PcT{C}{V_1}{V_2}$ and for all $V$,\\
$\PcT{\PUSH{C}{\ABST{x}{V}{}}}{T_1}{T_2}$.

\item ($\eta$-conversion for the terms that convert to $\lambda$-abstractions)
\objlabel{pc3_eta}

If $\PcT{C}{T}{\ABST{x}{W}{U}}$ and $\PcT{C}{V}{W}$ and $x \notin \FV{T}$
then\\ $\PcT{C}{\ABST{x}{V}{\APPL{x}{T}}}{T}$.

\end{enumerate}

\end{theorem}

\begin{proof}
Clause \thcref{}{pr0_subst0} is proved by induction on the first premise and
by cases on the second premise.
Clause \thcref{}{pr0_confluence} is proved by induction on $T_0$ and by cases on
the two premises. Here we must assume that the inductive hypothesis holds for
all proper subterms of $T_0$.
Clause \thcref{}{pr3_confluence} is a standard corollary of the previous clause,
proved using the ``strip lemma'' \cite{Brn92}.
Clauses \thcref{}{pc3_thin_dx_appl}, \thcref{}{pc3_head_1}, \thcref{}{pc3_head_2} are
immediate.
Clause \thcref{}{pc3_gen_abst} is proved by induction on the premise with the
standard technique used for generation lemmas \cite{Brn92}.
Clause \thcref{}{pc3_eta} is a corollary of clause \thcref{}{pc3_thin_dx_appl}.
\end{proof}

The main result on reduction is Church-Rosser property, while the main result
on conversion is its generation lemma on abstraction: a desirable property
mentioned in \cite{Dln80}.
The other properties, stating that conversion is a congruence, are referenced 
in \appref{mtt}.

What follows is a classification of the normal terms having an arity:

\begin{theorem}[the normal terms with an arity]\hfil
\objlabel{arity_nf2_inv_all}

If $\Arity{g}{C}{T}{L}$ and $\NfS{C}{T}$ then
there exist $\overline{V}$, $U$, $W$, $x$, $h$ such that:

\begin{enumerate}

\item
$T = \ABST{x}{W}{U}$ and
$\NfS{C}{W}$ and
$\NfS{\PUSH{C}{\ABST{x}{W}{}}}{U}$ or

\item
$T = \SORT{h}$ or

\item
$T = \APPL{\overline{V}}{\LREF{x}}$ and
$\NfsS{C}{\overline{V}}$ and
$\NfS{C}{\LREF{x}}$.

\end{enumerate}

\end{theorem}

\begin{proof}
By induction on the first premise and by cases on the second premise.
\end{proof}

The strong normalization theorem outlined below, stating that every term with
an arity is strongly normalizable, is one of the relevant results of the
present paper.

If we consider the connections between $\LD$ and $\LR$ that we briefly 
sketched in \subsecref{arity-defs}, it should not be a surprise that the
proof of strong normalization proposed by Tait for $\LR$ can be adapted for
$\LD$.
Namely both the definition of the strong reducibility candidates and the
overall proof method are the same.

Our formalization follows essentially the version of Tait's proof reported by
\cite{Loa98}. Other references we considered are \cite{LS04,GTL89,Ces01,Oos02}.
The main difference with respect to \cite{Loa98} is that we can use
abbreviations in place of explicit substitutions because of the shape of our
$\beta$-reductum (see \figref{pr0}{\beta}).

\begin{definition}[the strong reducibility candidates]\hfil
\objlabel{sc3}

The subset of the focalized terms that are strong reducibility candidates of
arity $L$ (with respect to the parameter $g$) is here denoted by
$\ScT{g}{L}{}{}$ and it is defined below.

\begin{footnotesize}
\[\left\{\begin{tabular}{lll}%
$\ScT{g}{\NODE{k}{h}}{E}{T}$&%
\hspace{-5pt}iff&\hspace{-5pt}%
$\Arity{g}{E}{T}{\NODE{k}{h}}$ and
$\SnT{E}{T}$
\\
$\ScT{g}{\IMPL{L_1}{L_2}}{E}{T}$&%
\hspace{-5pt}iff&\hspace{-5pt}%
$\Arity{g}{E}{T}{\IMPL{L_1}{L_2}}$ and
for each $C$, $C_1$, $C_2$, $V$,\\
&%
&\hspace{-5pt}%
$\ScT{g}{L_1}{C}{V}$ and $\PUSHT{C}{C_1}{E}{C_2}$
imply $\ScT{g}{L_2}{C}{\APPL{V}{T}}$
\end{tabular}\right.\]
\end{footnotesize}

\end{definition}

Notice that the possibility of exchanging the binders of the environment $C$
is silently assumed at least in \thcref{sn3_props}{sc3_sn3} below 
(see \cite{Loa98}).
Thus \defcref{sc3}{} must be rephrased carefully when binders are referenced
by position instead of by name (i.e with de Bruijn indexes) as in \cite{Gui05}
(see \defcref{sc3-real}{}).

We also define a version of the relaxed preorder on environments
(\defcref{csuba}{}) for use with the strong reducibility candidates, which we
need in \thcref{sn3_props}{sc3_arity_csubc}.

\begin{definition}[relaxed preorder on environments for candidates]\hfil
\objlabel{csubc}

The relation $\Csubc{g}{E_1}{E_2}$ is defined like $\Csuba{g}{E_1}{E_2}$ but
\defref{csuba}{abst} is replaced by the axiom below.
The notation ``rc'' stands for ``reducibility candidates''.

\begin{itemize}

\item (abst)
if $\Csubc{g}{C_1}{C_2}$ and
$\ScT{g}{L}{C_2}{V}$ and $\ScT{g}{\Asucc{g}{L}}{C_1}{W}$ then\\
$\Csubc{g}{\PUSH{C_1}{\ABST{x}{W}{}}}{\PUSH{C_2}{\ABBR{x}{V}{}}}$.

\end{itemize}

\end{definition}

Here are the main results on the preorder we just defined:

\begin{theorem}[main properties of the relation $\Csubc{g}{}{}$]\hfil
\objlabel{csubc_props}

\begin{enumerate}\EnumStyle

\item (the preorder for candidates implies the relaxed preorder)
\objlabel{csubc_csuba}

If $\Csubc{g}{C_1}{C_2}$ then $\Csuba{g}{C_1}{C_2}$.

\item (monotonicity of the arity assignment with respect to $\Csubc{g}{}{}$)
\objlabel{csubc_arity_conf}

If $\Csubc{g}{C_1}{C_2}$ and
$\Arity{g}{C_1}{T}{L}$ then
$\Arity{g}{C_2}{T}{L}$.

\end{enumerate}

\end{theorem}

\begin{proof}
Clause \thcref{}{csubc_csuba} is easily proved by induction on its premise.
Clause \thcref{}{csubc_arity_conf} is a corollary of the previous clause and of
\thcref{arity_props}{csuba_arity}.
\end{proof}

The strong normalization property, which we write as $\Arity{g}{C}{T}{L}$
implies $\SnT{g}{T}$, is not proved as is, but is derived from a number of
lemmas, which must be suitably generalized in order to be proved.

\begin{theorem}[main properties of the strongly normalizable terms]\hfil
\objlabel{sn3_props}

\begin{enumerate}\EnumStyle

\item (normal terms are strongly normalizable)
\objlabel{sn3_nf2}

If $\NfS{C}{T}$ then
$\SnT{C}{T}$.

\item (candidate type cast)
\objlabel{sc3_cast}

If $\ScT{g}{\Asucc{g}{L}}{C}{\APPL{\overline{V}}{V}}$ and
$\ScT{g}{L}{C}{\APPL{\overline{V}}{T}}$ then
$\ScT{g}{L}{C}{\APPL{\overline{V}}{\CAST{V}{T}}}$.

\item (candidate reference to abbreviation)
\objlabel{sc3_abbr}

If $\ScT{g}{L}{C}{\APPL{\overline{V}}{V}}$ and
$\GETL{C}{D}{\ABBR{x}{V}{D\p}}$ then
$\ScT{g}{L}{C}{\APPL{\overline{V}}{\LREF{x}}}$.

\item (candidate reference to abstraction)
\objlabel{sc3_abst}

If $\Arity{g}{C}{\APPL{\overline{V}}{\LREF{x}}}{L}$ and
$\NfS{C}{\LREF{x}}$ and
$\SnsT{C}{\overline{V}}$ then
$\ScT{g}{L}{C}{\APPL{\overline{V}}{\LREF{x}}}$.

\item (candidates are strongly normalizable)
\objlabel{sc3_sn3}

If $\ScT{g}{L}{C}{T}$ then
$\SnT{C}{T}$.

\item (candidate abbreviation)
\objlabel{sc3_bind}

If $\ScT{g}{L_2}{\PUSH{C}{\ABBR{x}{V}{}}}{\APPL{\overline{V}}{T}}$ and
$\ScT{g}{L_1}{C}{V}$ then
$\ScT{g}{L_2}{C}{\APPL{\overline{V}}{\ABBR{x}{V}{T}}}$.

\item (candidate $\beta$-redex)
\objlabel{sc3_appl}

If $\ScT{g}{L_2}{C}{\APPL{\overline{V}}{\ABBR{x}{V}{T}}}$ and
$\ScT{g}{L_1}{C}{V}$ and
$\ScT{g}{\Asucc{g}{L_1}}{C}{W}$ then
$\ScT{g}{L_2}{C}{\APPL{\overline{V}}{\APPL{V}{\ABST{x}{W}{T}}}}$.

\item (terms with an arity are candidates, general case)
\objlabel{sc3_arity_csubc}

If $\Arity{g}{C_1}{T}{L}$ and
$\GETL{E}{C_1}{D}$ and
$\Csubc{g}{E}{C_2}$ then
$\ScT{g}{L}{C_2}{T}$.

\item (terms with an arity are candidates)
\objlabel{sc3_arity}

If $\Arity{g}{C}{T}{L}$ then
$\ScT{g}{L}{C}{T}$.

\end{enumerate}

\end{theorem}

\begin{proof}
Clause \thcref{}{sn3_nf2} is immediate.
Clauses \thcref{}{sc3_cast}. \thcref{}{sc3_abbr}, \thcref{}{sc3_abst} and
\thcref{}{sc3_sn3} are proved by induction on $L$. 
Notice however that clauses \thcref{}{sc3_abst} and \thcref{}{sc3_sn3} must be
proved simultaneously.
Clauses \thcref{}{sc3_bind} and \thcref{}{sc3_appl} are proved by induction on $L_2$
by invoking clause \thcref{}{sc3_sn3}.
Clause \thcref{}{sc3_arity_csubc} is proved by induction on its first premise
and by cases on its third premise; here we invoke the clauses 
\thcref{}{sc3_cast}, \thcref{}{sc3_abbr}, \thcref{}{sc3_abst}, \thcref{}{sc3_bind},
\thcref{}{sc3_appl} with $\overline V$ as the empty list $\NIL$ but this
assumption is too weak to prove the clauses themselves; in the proof we also
invoke \thcref{csubc_props}{csubc_arity_conf}.
Clause \thcref{}{sc3_arity} follows from the previous clause.
\end{proof}

The fact that every term with an arity is strongly normalizing follows from
the composition of \thcref{sn3_props}{sc3_arity} (the main result) and
\thcref{sn3_props}{sc3_sn3}, but notice that the converse is not true in
general as we imply from \thcref{sn3_props}{sn3_nf2} and \thcref{nf2_ex2}{}.

%% file: native_props.tex
\subsection{Results on the Native Type Assignment}
\subseclabel{native-props}

The first result about the type system is the generation (i.e. inversion)
lemma, whose aim is to invert the type assignment rules of \defcref{ty3}{}.

\begin{theorem}[generation lemma for native type assignment]\hfil
\objlabel{ty3_gen}

\begin{enumerate}\EnumStyle

\item (generation lemma on sorts)
\objlabel{ty3_gen_sort}

If $\TyT{g}{C}{\SORT{h}}{T}$ then
$\PcT{C}{\SORT{\Next{g}{h}}}{T}$.

\item (generation lemma on bound references)
\objlabel{ty3_gen_lref}

If $\TyT{g}{C}{\LREF{x}}{T}$ then
there exist $E$, $E\p$, $V$, $U$ such that
$\PcT{C}{U}{T}$ and \\
$\GETL{C}{E}{\ABBR{x}{V}{E\p}}$ and
$\TyT{g}{E}{V}{U}$ or
there exist $E$, $E\p$, $V$, $U$ such that \\
$\PcT{C}{V}{T}$ and
$\GETL{C}{E}{\ABST{x}{V}{E\p}}$ and
$\TyT{g}{E}{V}{U}$.

\item (generation lemma on abbreviations)
\objlabel{ty3_gen_bind_abbr}

If $\TyT{g}{C}{\ABBR{x}{V}{U_1}}{T}$ then
there exist $U_2$, $U$ such that
$\PcT{C}{\ABBR{x}{V}{U_2}}{T}$ and
$\TyT{g}{C}{V}{U}$ and
$\TyT{g}{\PUSH{C}{\ABBR{x}{V}{}}}{U_1}{U_2}$.

\item (generation lemma on abstractions)
\objlabel{ty3_gen_bind_abst}

If $\TyT{g}{C}{\ABST{x}{V}{U_1}}{T}$ then
there exist $U_2$, $U$ such that
$\PcT{C}{\ABST{x}{V}{U_2}}{T}$ and
$\TyT{g}{C}{V}{U}$ and
$\TyT{g}{\PUSH{C}{\ABST{x}{V}{}}}{U_1}{U_2}$.

\item (generation lemma on applications)
\objlabel{ty3_gen_appl}

If $\TyT{g}{C}{\APPL{V_1}{U_1}}{T}$ then
there exist $V_2$, $U_2$ such that
$\PcT{C}{\APPL{V_1}{\ABST{x}{V_2}{U_2}}}{T}$ and
$\TyT{g}{C}{U_1}{\ABST{x}{V_2}{U_2}}$ and
$\TyT{g}{C}{V_1}{V_2}$.

\item (generation lemma on type annotations)
\objlabel{ty3_gen_cast}

If $\TyT{g}{C}{\CAST{V}{U}}{T}$ then
there exists $V_0$ such that
$\PcT{C}{\CAST{V_0}{V}}{T}$ and\\
$\TyT{g}{C}{U}{V}$ and
$\TyT{g}{C}{V}{V_0}$.

\end{enumerate}

\end{theorem}

\begin{proof}
All clauses are proved by induction on the premise with the standard technique
used to prove generation lemmas in general \cite{Brn92}.
\end{proof}

Some important properties of the native type assignment are listed below.

\begin{theorem}[main properties of native type assignment]\hfil
\objlabel{ty3_props}

\begin{enumerate}\EnumStyle

\item (thinning preserves type)
\objlabel{ty3_lift}

If $\TyT{g}{C_2}{T_1}{T_2}$ and
$\PUSHT{C_1}{D\p}{C_2}{D\p\p}$ then
$\TyT{g}{C_1}{T_1}{T_2}$.

\item (correctness of types)
\objlabel{ty3_correct}

If $\TyT{g}{C}{T_1}{T_2}$ then
there exists $T_3$ such that
$\TyT{g}{C}{T_2}{T_3}$.

\item (uniqueness of types up to conversion)
\objlabel{ty3_unique}

If $\TyT{g}{C}{T}{T_1}$ and
$\TyT{g}{C}{T}{T_2}$ then
$\PcT{C}{T_1}{T_2}$.

\item (substitution in focalized terms preserves the type)
\objlabel{ty3_fsubst0}

If $\TyT{g}{C_1}{T_1}{T}$ and
$\FsubstZ{x}{V}{C_1}{T_1}{C_2}{T_2}$ and
$\GETL{C_1}{E}{\ABBR{x}{V}{E\p}}$ then
$\TyT{g}{C_2}{T_2}{T}$.

\item (substitution in terms preserves the type)
\objlabel{ty3_subst0}

If $\TyT{g}{C}{T_1}{T}$ and
$\SubstZ{x}{V}{T_1}{T_2}$ and
$\GETL{C}{E}{\ABBR{x}{V}{E\p}}$ then
$\TyT{g}{C}{T_2}{T}$.

\item (substitution in environments preserves the type)
\objlabel{ty3_csubst0}

If $\TyT{g}{C_1}{T}{T_0}$ and
$\CsubstZ{x}{V}{C_1}{C_2}$ and
$\GETL{C_1}{E}{\ABBR{x}{V}{E\p}}$ then
$\TyT{g}{C_2}{T}{T_0}$.

\item (monotonicity of the type assignment with respect to $\Csubt{g}{}{}$)
\objlabel{csubt_ty3}

If $\TyT{g}{C_1}{T_1}{T_2}$ and
$\Csubt{g}{C_1}{C_2}$ then
$\TyT{g}{C_2}{T_1}{T_2}$.

\item (type checking implies type inference)
\objlabel{ty3_typecheck}

If $\TyT{g}{C}{T}{V}$ then
there exists $U$ such that
$\TyT{g}{C}{\CAST{V}{T}}{U}$.

\end{enumerate}

\end{theorem}

\begin{proof}
Clause \thcref{}{ty3_lift} is proved by induction on the first premise.
The proof of clauses \thcref{}{ty3_correct} and \thcref{}{ty3_unique} is by
induction on their first premise and contains invocations of
\thcref{ty3_gen}{} and of clause \thcref{}{ty3_lift}.
Clause \thcref{}{ty3_fsubst0} is proved by double induction on the first two
premises and by invoking the previous clauses.
The statements \thcref{}{ty3_subst0} and \thcref{}{ty3_csubst0} are mutually
recursive so we prove them as corollaries of clause \thcref{}{ty3_fsubst0}.
Clause \thcref{}{csubt_ty3} is proved by induction on the first premise.
Clause \thcref{}{ty3_typecheck} is a corollary of clause \thcref{}{ty3_correct}. 
\end{proof}

A consequence of \thcref{ty3_gen}{ty3_gen_cast} is that if $\CAST{W}{T}$ is
typable in $E$ then $T$ has type $W$ in $E$.
The converse also holds by \thcref{ty3_props}{ty3_typecheck} and this implies
that in $\LD$, type checking can be expressed in terms of type inference
\cite{Brn92}.

\thcref{ty3_props}{csubt_ty3} is the most relevant result about the
preorder $\Csubt{g}{}{}$.

The subject reduction of $\LD$ is one of the main results we are presenting
in this paper. The main part of the proof is concentrated in the base case,
where a single step of environment-free parallel reduction is considered.
The possibility to reduce some terms appearing inside the environment is
essential here (see \cite{KBN99}).
The general case is just a simple corollary.

\begin{theorem}[subject reduction and corollaries]\hfil
\objlabel{ty3_sred}

\begin{enumerate}\EnumStyle

\item (base case)
\objlabel{ty3_sred_wcpr0_pr0}

If $\TyT{g}{C_1}{T}{T_2}$ and
$\WcprZ{C_1}{C_2}$ and
$\PrZ{T}{T_1}$ then
$\TyT{g}{C_2}{T_1}{T_2}$.

\item (general case without the reduction in the environment)
\objlabel{ty3_sred_pr3}

If $\PrT{C}{T}{T_1}$ and
$\TyT{g}{C}{T}{T_2}$ then
$\TyT{g}{C}{T_1}{T_2}$.

\item (inverse of type preservation by thinning)
\objlabel{ty3_gen_lift}

If $\TyT{g}{C_1}{T}{T_1}$ and
$\PUSHT{C_1}{D\p}{C_2}{D\p\p}$ then
there exists $T_2$ such that \\
$\PcT{C_1}{T_2}{T_1}$ and
$\TyT{g}{C_2}{T}{T_2}$.

\item (type reduction)
\objlabel{ty3_tred}

If $\TyT{g}{C}{T}{T_1}$ and
$\PrT{C}{T_1}{T_2}$ then
$\TyT{g}{C}{T}{T_2}$.

\item (subject conversion: first case)
\objlabel{ty3_sconv_pc3}

If $\TyT{g}{C}{U_1}{T_1}$ and
$\TyT{g}{C}{U_2}{T_2}$ and
$\PcT{C}{U_1}{U_2}$ then
$\PcT{C}{T_1}{T_2}$.

\item (subject conversion: second case)
\objlabel{ty3_sconv}

If $\TyT{g}{C}{U_1}{T_1}$ and
$\TyT{g}{C}{U_2}{T_2}$ and
$\PcT{C}{U_1}{U_2}$ then
$\TyT{g}{C}{U_1}{T_2}$.

\end{enumerate}

\end{theorem}

\begin{proof}
Clause \thcref{}{ty3_sred_wcpr0_pr0} is proved by induction on the first premise
and by cases on the third premise with frequent invocations of
\thcref{ty3_gen}{}, \thcref{ty3_props}{ty3_lift} and
\thcref{ty3_props}{ty3_correct}.
In the case of \figref{ty3}{abbr} against \figref{pr0}{\delta} we exploit 
\thcref{ty3_props}{ty3_subst0}, and in the case of \figref{ty3}{appl} against
\figref{pr0}{\beta} we exploit \thcref{ty3_props}{csubt_ty3}.
Clause \thcref{}{ty3_sred_pr3} is corollary of the previous clause proved by
induction on the first premise.
Clause \thcref{}{ty3_gen_lift} is proved by induction on the first premise.
Clauses \thcref{}{ty3_tred} and \thcref{}{ty3_sconv} are corollaries of
\thcref{ty3_props}{ty3_correct}.
Clause \thcref{}{ty3_sconv_pc3} is a corollary of \thcref{ty3_props}{ty3_unique}.
\end{proof}

We would like to stress that the proof of the subject reduction is more
difficult in $\LD$ than in the $\lambda$-cube because in $\LD$ we can not
assume that the type of the type of a term is a sort (as it is often done in
$\lambda$-cube).

With \thcref{ty3_sred}{ty3_sred_wcpr0_pr0} we avoid the simultaneous induction
with which many authors, including \cite{KBN99}, prove the results like
\thcref{ty3_sred}{ty3_sred_pr3}.
Notice that \thcref{ty3_sred}{ty3_sconv} is stated as a desired property in
\cite{Dln80}.

Some properties of the type system are proved more easily invoking arities
because arities are assigned up to level equality instead of up to conversion
and level equality is easier to manage being defined by simpler rules. 
The other rules of the arity assignment have the same complexity of the
corresponding rule for the types.

\begin{theorem}[some properties of types proved using arities]\hfil
\objlabel{ty3_arity_props}

\begin{enumerate}\EnumStyle

\item (typed terms have an arity)
\objlabel{ty3_arity}

If $\TyT{g}{C}{T_1}{T_2}$ then
there exists $L$ such that
$\Arity{g}{C}{T_1}{L}$ and
$\Arity{g}{C}{T_2}{\Asucc{g}{L}}$.

\item (typed terms are strongly normalizable)
\objlabel{ty3_sn3}

If $\TyT{g}{C}{T}{U}$ then
$\SnT{C}{T}$.

\item (the preorder on environments implies the relaxed preorder)
\objlabel{csubt_csuba}

If $\Csubt{g}{C_1}{C_2}$ then $\Csuba{g}{C_1}{C_2}$.

\item (abstraction is predicative)
\objlabel{ty3_predicative}

If $\TyT{g}{C}{\ABST{x}{V}{T}}{U}$ then
$\NotPcT{C}{U}{V}$.

\item (abstraction is not absorbent)
\objlabel{ty3_repellent}

If $\TyT{g}{C}{\ABST{x}{V}{T}}{U_1}$ and
$\TyT{g}{\PUSH{C}{\ABST{x}{V}{}}}{T}{U_2}$ and
$x \notin \FV{U_2}$ then
$\NotPcT{C}{U_1}{U_2}$.

\item (terms can not be typed with themselves)
\objlabel{ty3_acyclic}

If $\TyT{g}{C}{T}{U}$ then $\NotPcT{C}{U}{T}$.

\end{enumerate}

\end{theorem}

\begin{proof}
Clause \thcref{}{ty3_arity} is a consequence of \thcref{arity_sred}{}, it is
proved by induction on its premise and it is a prerequisite of the other
clauses.
In particular clause \thcref{}{ty3_sn3} is a corollary of
\thcref{sn3_props}{sc3_arity} and \thcref{sn3_props}{sc3_sn3},
Clause \thcref{}{csubt_csuba} is proved by induction on its premise by invoking
\thcref{arity_props}{csuba_arity}.
clause \thcref{}{ty3_predicative} invokes \thcref{ty3_gen}{ty3_gen_bind_abst},
clause \thcref{}{ty3_repellent} invokes \thcref{ty3_props}{ty3_correct}, and
clause \thcref{}{ty3_acyclic} uses the strict monotonicity condition of the sort
hierarchy parameter $g$ (see \defcref{G}{}).
\end{proof}

Notice that \thcref{ty3_arity_props}{ty3_arity} includes our version of the
theorem stating that the level of a term and the level of its type differ in
one application of the successor function (originally proved by de Bruijn for
his calculi). 

\thcref{ty3_arity_props}{ty3_predicative} states that a term
constructed by abstraction never belongs to the abstraction domain (i.e.
the class of the terms typed by $V$ in this case).
Moreover \thcref{ty3_arity_props}{ty3_repellent} states that
in $\LD$ there is no term $*$ for which, in standard notation:
\[
\Rule
 {\Gamma \vdash A \mathrel: * \qquad
  \Gamma, x\mathord:A \vdash B \mathrel: *}{}
 {\Gamma \vdash (\lambda_{x:A}.B) \mathrel: *}
\]
We stress that \thcref{ty3_arity_props}{ty3_predicative} and
\thcref{ty3_arity_props}{ty3_repellent} are expected properties of the
$\lambda$-abstraction, which hold in every typed $\lambda$-calculus.

The decidability results we present below are a consequence of
\thcref{ty3_arity_props}{ty3_sn3}.

\begin{theorem}[main decidability results]\hfil
\objlabel{ty3_dec}

\begin{enumerate}\EnumStyle

\item (convertibility of typed terms is decidable)
\objlabel{pc3_dec}

If $\TyT{g}{C}{U_1}{T_1}$ and
$\TyT{g}{C}{U_2}{T_2}$ then
$\PcT{C}{U_1}{U_2}$ or
$\NotPcT{C}{U_1}{U_2}$.

\item (type inference is decidable)
\objlabel{ty3_inference}

For all $C$, $T_1$ there exists $T_2$ such that $\TyT{g}{C}{T_1}{T_2}$
or for all $T_2$, $\NotTyT{g}{C}{T_1}{T_2}$.

\end{enumerate}

\end{theorem}

\begin{proof}
Clause \thcref{}{pc3_dec} is a standard consequence of
\thcref{ty3_arity_props}{ty3_sn3} and \thcref{prc_props}{pr3_confluence}.
Clause \thcref{}{ty3_inference} is proved by induction on the focalized term
$(C, T_1)$ using \thcref{ty3_gen}{}, \thcref{ty3_props}{ty3_correct}, 
\thcref{ty3_sred}{ty3_sred_pr3} and the previous clause.
We assume that the inductive hypothesis holds for all proper subterms of
$(C, T_1)$ (intended as the term $C.T_1$).
Moreover we consider $(\PUSH{E}{\ABST{x}{W}{}}, T)$ and
$(\PUSH{E}{\ABBR{x}{V}{}}, T)$ as subterms of $(E, \ABST{x}{W}{T})$ and
$(E, \ABBR{x}{V}{T})$ respectively (because of \figref{ty3}{abst}
and \figref{ty3}{abbr}).
\end{proof}

Notice that by 
\thcref{ty3_gen}{ty3_gen_cast} and \thcref{ty3_props}{ty3_typecheck},
type checking is also decidable.

%% file: static_props.tex
\subsection{Results on the Static Type Assignment}
\subseclabel{static-props}

The main results about $\StyZ{g}{C}{}{}$ are listed below.

\begin{theorem}[main properties of the static type]\hfil
\objlabel{sty_props}

\begin{enumerate}\EnumStyle

\item (a typable term is typed by its static type)
\objlabel{ty3_sty0}

If $\TyT{g}{C}{U}{T_1}$ and
$\StyZ{g}{C}{U}{T_2}$ then
$\TyT{g}{C}{U}{T_2}$.

\item (the iterated static type yields a term that can be seen as an environment)
\objlabel{sty1_env}

If $\StyZ{g}{C}{T_1}{T}$ then
there exists $T_2$ such that
$\StyF{g}{C}{T_1}{T_2}$ and
$\Env{T_2}$.

\end{enumerate}

\end{theorem}

\begin{proof}
Clause \thcref{}{ty3_sty0} is proved by induction on the first premise and by
cases on the second premise.
While considering \figref{ty3}{appl} and \figref{ty3}{cast}, we invoke
\thcref{ty3_gen}{}, \thcref{ty3_props}{ty3_correct},
\thcref{ty3_props}{ty3_unique} and \thcref{ty3_sred}{ty3_sconv}.
Clause \thcref{}{sty1_env} is easily proved by induction on its premise.
\end{proof}

\thcref{sty_props}{ty3_sty0} shows that the static type is indeed a type if
we compute it on typed (i.e. legal) terms, and we can consider it as the
canonical type of that term in the sense of \cite{KN96a}.

\thcref{sty_props}{sty1_env} allows to map a term $T$ to the environment
$\gamma(T)$ obtained iterating the static type assignment on $T$ the least
number of times.
Once extended arbitrarily on not well typed terms, $\gamma$ yields an
immersion of $\TRM$ into $\ENV$.
The above considerations clearly justify the choice of the function
$\StyF{g}{E}{}{}$ as the main ingredient for switching between terms and
environments in the $\LD$ setting. 
Notice that $\gamma$ and its properties have not being formally specified yet
because the behavior of this function, especially with respect to reduction,
is expected to be much clearer when the duality between terms and contexts
will be achieved (see \appref{duality}).

%% file: examples.tex
\subsection{Examples}
\subseclabel{examples}

If we consider the concrete sort hierarchy parameter $\Gz$ defined by
$\Next{\Gz}{h} \equiv h + 1$, we have
$\Leq{\Gz}{\NODE{k_1}{h_1}}{\NODE{k_2}{h_2}}$ iff $h_1 + k_2 = h_2 + k_1$
and we know that $\NAT \times \NAT$ (i.e. the set of the nodes) equipped with
this equality is isomorphic to the set of the integer numbers.
To formalize this assertion, we define the integer level equality on nodes, we
extend it on compound arities, and we state the following theorem.

\begin{definition}[integer level quality]\hfil
\objlabel{leqz}

The integer level equality predicate $\Leqz{L_1}{L_2}$ is defined by the rules
in \figref{leqz}{}.

\end{definition}

\begin{figure}
\begin{center}

\Rule
 {k_1 + h_2 = k_2 + h_1}{\mathrm{node}}
 {\Leqz{\NODE{k_1}{h_1}}{\NODE{k_2}{h_2}}} \quad
\Rule
 {\Leqz{L_1}{L_2} \quad \Leqz{L_3}{L_4}}{\mathrm{impl}}
 {\Leqz{\IMPL{L_1}{L_3}}{\IMPL{L_2}{L_4}}}

\end{center}
\caption{Integer level equality rules}
\figlabel{leqz}
\end{figure}

\begin{theorem}[level equality for the concrete parameter $\Gz$]\hfil
\objlabel{levels_ex0}

\begin{enumerate}\EnumStyle

\item (level equality for $\Gz$ implies integer level equality)
\objlabel{leqz_leq}

If $\Leq{gz}{L_1}{L_2}$ then
$\Leqz{L_1}{L_2}$.

\item (integer level equality implies level equality for $\Gz$)
\objlabel{leq_leqz}

If $\Leqz{L_1}{L_2}$ then
$\Leq{gz}{L_1}{L_2}$.

\end{enumerate}

\end{theorem}

\begin{proof}
Both clauses are easily proved by induction on their premises.
\end{proof}

The converse of \thcref{ty3_arity_props}{ty3_arity} is not true in general in
fact there are terms that have an arity but that are not typable. The next
result shows an example.

\begin{theorem}[an untypable term having an arity]\hfil
\objlabel{ty3_ex1}

Given the term $T \equiv \APPL{\LREF{x_2}}{\ABST{x_3}{\LREF{x_0}}{\SORT{0}}}$
in the environment 

$E \equiv \ABST{x_0}{\SORT{0}}{\ABST{x_1}{\SORT{0}}{\ABST{x_2}{\LREF{x_1}}{\SORT{0}}}}$
we have that: 

\begin{enumerate}\EnumStyle

\item ($T$ has an arity in $E$)
\objlabel{ex1_arity}

$\Arity{g}{E}{T}{\NODE{0}{0}}$.

\item ($T$ is not typable in $E$)
\objlabel{ex1_ty3}

For all $U$, $\NotTyT{g}{E}{T}{U}$.

\end{enumerate}

\end{theorem}

\begin{proof}
Clause \thcref{}{ex1_arity} is immediate.
Clause \thcref{}{ex1_ty3} is a consequence of \thcref{ty3_gen}{}.
\end{proof}

The next theorem shows that there are normal terms that do not have an arity.

\begin{theorem}[a normal term without an arity]\hfil
\objlabel{nf2_ex2}

Given the term $T \equiv \APPL{\SORT{0}}{\SORT{0}}$
in the environment $E \equiv \SORT{0}$,
we have that: 

\begin{enumerate}\EnumStyle

\item ($T$ is normal in $E$)
\objlabel{ex2_nf2}

$\NfS{E}{T}$.

\item ($T$ does not have an arity in $E$)
\objlabel{ex2_arity}

For all $L$, $\NotArity{g}{E}{T}{L}$.

\end{enumerate}

\end{theorem}

\begin{proof}
Both clauses are immediate consequences of simple generation lemmas, which we
prove by induction on the premise with a standard technique.
\end{proof}

%% file: extension.tex
\section{The Extension of $\LD$ with the Exclusion Binder $\chi$}
\seclabel{extension}

In this section we present the calculus $\CLD$ by which we mean the calculus
$\LD$ extended by adding the exclusion binder $\VOID{}{}$
(see \subsecref{exclusion}).
In this extension we show that every environment has a canonical well-formed
form in the usual sense (see \subsecref{legal}), which preserves the native
type assignment.

%% file: exclusion.tex
\subsection{The Calculus $\CLD$}
\subseclabel{exclusion}

In this subsection we extend $\LD$ by adding the \emph{exclusion binder}
that here we call $\VOID{}{}$ 
(after $\chi \acute{\alpha} o \sigma $: Greek for ``gaping void'').
The calculus we obtain is called $\CLD$ and is the one we formalized in
\cite{Gui05}.
The idea behind the exclusion binder is that a variable $x$ bound by
$\VOID{x}{}$ is \emph{excluded} in the sense that it must not occur in the
scope of $\VOID{x}{}$.
The intended use of this binder is to replace the other binders of an
environment when they are not referenced. In this way we erase these
binders from the environment without changing its length.
This binder-erasing technique is particularly efficient when the bound
variables are referenced by position (i.e using the so-called \emph{de Bruijn
indexes} \cite{SPAc2}) instead of by name.  

\begin{definition}[exclusion item]\hfil

We introduce the syntactic item $\VOID{x}{}$ (exclusion) and we extend the
syntax of terms and environments as follows:
\[ \TRM \equiv \TRM \| \VOID{\VAR}{\TRM} \qquad
   \ENV \equiv \ENV \| \VOID{\VAR}{\ENV}
\]
\end{definition}

The construction $\VOID{x}{T}$ ($\chi$-abstraction) is thought as well formed
if $x \notin \FV{T}$.

We want the $\VOID{}{}$ binder to have the reductional behavior of the
unreferenced abbreviation, so we add the $\zeta$-contraction and the 
$\upsilon$-swap of \figref{environment-free-void}{}.

\begin{figure}
\begin{center}

\begin{tabular}{|l|llll|}
\hline
\textbf{scheme} & \textbf{redex} & & \textbf{reductum} & \\ \hline

$\zeta$-contraction &
$\VOID{x}{T}$ & $\SR{\zeta}$ & $T$ & if $x \notin \FV{T}$ \\ \hline

$\upsilon$-swap &
$\APPL{V}{\VOID{x}{T}}$ & $\SR{\upsilon}$ &
$\VOID{x}{\APPL{V}{T}}$ & \\ \hline

\end{tabular}

\end{center}
\caption{Environment-free reduction steps for exclusion}
\figlabel{environment-free-void}
\end{figure}

Formally we obtain this behavior by adding the rules of \figref{pr0-void}{}.

\begin{figure}
\begin{center}
\Rule{\PrZ{T_1}{T_2}}{\mathrm{comp}}
     {\PrZ{\VOID{x}{T_1}}{\VOID{x}{T_2}}} \quad
\Rule
 {\PrZ{T_1}{T_2} \quad x \notin \FV{T_1}}{\zeta}
 {\PrZ{\VOID{x}{T_1}}{T_2}} \quad
\Rule
 {\PrZ{V_1}{V_2} \quad \PrZ{T_1}{T_2}}{\upsilon}
 {\PrZ{\APPL{V_1}{\VOID{x}{T_1}}}{\VOID{x}{\APPL{V_2}{T_2}}}}
\end{center}
\caption{Reduction rules for the exclusion binder}
\figlabel{pr0-void}
\end{figure}

The general type assignment policy of the $\chi$-abstraction follows that of
the abbreviation but we do not add a rule for typing an excluded variable
occurrence. In this way we capture our intuition of the exclusion because the
excluded variable occurrences remain untyped.
This policy applies uniformly to the assignment of the native type, of the
static type and of the arity as we see in \figref{types-void}{}. 

\begin{figure}
\begin{center}
\Rule
 {\TyT{g}{\PUSH{C}{\VOID{x}{}}}{T}{U}}{\mathrm{void}}
 {\TyT{g}{C}{\VOID{x}{T}}{\VOID{x}{U}}} \quad
\Rule
 {\StyZ{g}{\PUSH{C}{\VOID{x}{}}}{T}{U}}{\mathrm{void}}
 {\StyZ{g}{C}{\VOID{x}{T}}{\VOID{x}{U}}} \quad
\Rule
 {\Arity{g}{\PUSH{C}{\VOID{x}{}}}{T}{L}}{\mathrm{void}}
 {\Arity{g}{C}{\VOID{x}{T}}{L}}
\end{center}
\caption{Typing rules for the exclusion binder}
\figlabel{types-void}
\end{figure}

The domain-based preorder on environments is extended by defining the domain
of an excluded variable occurrence $x$ as the whole set $\TRM$ of terms
because being never well formed, $x$ can be a placeholder for any term.

\begin{definition}[preorders on environments for exclusion]\hfil

Under the assumption $\Csubt{g}{E_1}{E_2}$ we set 
$\Csubt{g}{\PUSH{E_1}{\VOID{x}{}}}{\PUSH{E_2}{\VOID{x}{}}}$ and 
$\Csubt{g}{\PUSH{E_1}{\VOID{x}{}}}{\PUSH{E_2}{\ABST{x}{W}{}}}$ and 
$\Csubt{g}{\PUSH{E_1}{\VOID{x}{}}}{\PUSH{E_2}{\ABBR{x}{V}{}}}$.
We do the same for the preorders $\Csuba{g}{}{}$ and $\Csubc{g}{}{}$.

\end{definition}

We also need the rules stating the compatibility of the $\chi$-abstraction
with the context predicate (\defcref{env}{}), with the substitution
(\defcref{subst0}{}, \defcref{csubst0}{}) and with the weak reduction of
environments (\defcref{wcpr0}{}).

Every theorem we stated $\LD$ holds in $\CLD$ as well, in addition we can
prove:

\begin{theorem}[main properties of exclusion]\hfil
\objlabel{void_props}

\begin{enumerate}\EnumStyle

\item (compatibility with environment-dependent parallel conversion)
\objlabel{pc3_head_2_void}

If $\PcT{\PUSH{C}{\VOID{x}{}}}{T_1}{T_2}$ then
$\PcT{C}{\VOID{x}{T_1}}{\VOID{x}{T_2}}$.

\item (candidate exclusion)
\objlabel{sc3_void}

If $\ScT{g}{L_2}{\PUSH{C}{\VOID{x}{}}}{\APPL{\overline{V}}{T}}$ then
$\ScT{g}{L_2}{C}{\APPL{\overline{V}}{\VOID{x}{T}}}$.

\item (generation lemma for native type assignment)
\objlabel{ty3_gen_void}

If $\TyT{g}{C}{\VOID{x}{U_1}}{T}$ then
there exists $U_2$ such that
$\PcT{C}{\VOID{x}{U_2}}{T}$ and \\
$\TyT{g}{\PUSH{C}{\VOID{x}{}}}{U_1}{U_2}$.

\end{enumerate}

\end{theorem}

\begin{proof}
Clause \thcref{}{pc3_head_2_void} is proved like \thcref{prc_props}{pc3_head_2}.
Clause \thcref{}{sc3_void} is proved like \thcref{sn3_props}{sc3_bind}.
Clause \thcref{}{ty3_gen_void} is proved like \thcref{ty3_gen}{ty3_gen_bind_abbr}.
\end{proof}

%% file: legal.tex
\subsection{Legal Environments in $\CLD$}
\subseclabel{legal}

In some versions of the $\lambda$-cube \cite{KBN99} and in other type
theories \cite{MS05}, the rule for typing a variable declared in an
environment (the so-called ``start'' rule) requires that the environment is
legal (or well formed), which means that every declaration or definition in
the environment is well typed.
Following \cite{Brn92}, in \subsecref{native-defs} we showed that the explicit
notion of a legal environment is not essential for defining our type judgement.
However we may be interested in this notion for several reasons. 
For instance in the set theoretic semantics of a $\lambda$-calculus
\cite{Jac99}, a term typed in an environment is denoted (approximately) by a
function taking an argument for each environment entry, thus all the
environment entries must be typable.

In this section we use the exclusion binder $\VOID{}{}$ to define the
``default legal version'' of an arbitrary $\CLD$-environment (that in
particular can be a $\LD$-environment), and we show that the type of a term is
preserved when we ``legalize'' the environment.

Given an environment $E$, we introduce its default legal form $\WfT{g}{E}{}$
(the abbreviation of ``well formed'' taken from \cite{Coq}) that is $E$ with
the non-binding entries removed and with the untypable entries replaced by
$\chi$.

By using the $\chi$ binder, the environment $\WfT{g}{E}{}$ has the length of
the environment $E$ and the terms referring to $E$ can refer to $\WfT{g}{E}{}$
without being relocated.
This feature is desired in the formal specification of $\CLD$ \cite{Gui05},
where the environment entries are referred by position, and not by name as in
this paper.

Notice that the function $\WfT{g}{}{}$ is well defined and total because
the type inference problem is decidable in $\CLD$
(see \thcref{ty3_dec}{ty3_inference}).
Also notice that $\WfT{g}{}{}$ depends on the sort hierarchy parameter $g$
defined in \subsecref{native-defs}.

In \cite{Gui05} we do not have the function for inferring the type of a term,
therefore we prefer to define $\WfT{g}{}{}$ by axiomatizing the proposition
$\WfT{g}{E_1}{E_2}$.

\begin{definition}[environment legalization]\hfil 
\objlabel{wf3}

The default legalization of the environment $E$ is the environment
$\WfT{g}{E}{}$ defined by axiomatizing the predicate $\WfT{g}{E_1}{E_2}$ with
the following clauses:

\begin{enumerate}
\EnumStyle

\item \objlabel{wf3_sort}
$\WfT{g}{\SORT{h}}{\SORT{h}}$.

\item \objlabel{wf3_abbr_t}
If $\WfT{g}{E_1}{E_2}$ and
$\TyT{g}{E_1}{V}{W}$ then
$\WfT{g}{\PUSH{E_1}{\ABBR{x}{V}{}}}{\PUSH{E_2}{\ABBR{x}{V}{}}}$.

\item \objlabel{wf3_abst_t}
If $\WfT{g}{E_1}{E_2}$ and
$\TyT{g}{E_1}{W}{V}$ then
$\WfT{g}{\PUSH{E_1}{\ABST{x}{W}{}}}{\PUSH{E_2}{\ABST{x}{W}{}}}$.

\item \objlabel{wf3_void}
If $\WfT{g}{E_1}{E_2}$ then
$\WfT{g}{\PUSH{E_1}{\VOID{x}{}}}{\PUSH{E_2}{\VOID{x}{}}}$.

\item \objlabel{wf3_abbr_u}
If $\WfT{g}{E_1}{E_2}$ and
for each $W$, 
$\NotTyT{g}{E_1}{V}{W}$, then
$\WfT{g}{\PUSH{E_1}{\ABBR{x}{V}{}}}{\PUSH{E_2}{\VOID{x}{}}}$.

\item \objlabel{wf3_abst_u}
If $\WfT{g}{E_1}{E_2}$ and
for each $V$, 
$\NotTyT{g}{E_1}{W}{V}$, then
$\WfT{g}{\PUSH{E_1}{\ABST{x}{W}{}}}{\PUSH{E_2}{\VOID{x}{}}}$.

\item \objlabel{wf3_appl}
If $\WfT{g}{E_1}{E_2}$ then
$\WfT{g}{\PUSH{E_1}{\APPL{V}{}}}{E_2}$.

\item \objlabel{wf3_cast}
If $\WfT{g}{E_1}{E_2}$ then
$\WfT{g}{\PUSH{E_1}{\CAST{W}{}}}{E_2}$.

\end{enumerate}

\end{definition}

We do not give these axioms as rules because Axiom \thcref{}{wf3_abbr_u} and
Axiom \thcref{}{wf3_abst_u} are expressed in the meta-language and can not be
given in rule form.

The most relevant properties of the function  $\WfT{g}{}{}$ are listed in
the theorem below:

\begin{theorem}[main properties of the legalization function]\hfil
\objlabel{wf3_props}

\begin{enumerate}
\EnumStyle

\item (the legalization function is total)
\objlabel{wf3_total}

For all $C_1$, there exists $C_2$ such that $\WfT{g}{C_1}{C_2}$.

\item (preservation of the native type assignment)
\objlabel{wf3_ty3_conf}

If $\TyT{g}{C_1}{T}{U}$ and
$\WfT{g}{C_1}{C_2}$ then
$\TyT{g}{C_2}{T}{U}$.

\item (environments in native type assignments can be assumed legal)
\objlabel{wf3_ty3}

If $\TyT{g}{C_1}{T}{U}$ then
there exists $C_2$ such that
$\WfT{g}{C_1}{C_2}$ and
$\TyT{g}{C_2}{T}{U}$.

\end{enumerate}

\end{theorem}

\begin{proof}
Clause \thcref{}{wf3_total} is proved by induction on $C_1$ with the help of
\thcref{ty3_dec}{ty3_inference}.
Clause \thcref{}{wf3_ty3_conf} is proved by induction on its first premise; here
we need \thcref{ty3_props}{ty3_correct}, \thcref{ty3_sred}{ty3_sred_pr3} and
\thcref{ty3_dec}{ty3_inference}.
Clause \thcref{}{wf3_ty3} is implied the previous clauses.
\end{proof}

$\thcref{wf3_props}{wf3_ty3_conf}$ and $\thcref{wf3_props}{wf3_ty3}$ imply
each other but we noticed that the second one is slightly harder to prove
directly because its conclusion is existential.

%% file: conclusions.tex
\section{Conclusions and Future Work}
\seclabel{conclusions}

In this paper we take the calculus $\LI$ \cite{SPAc6} with the restricted
applicability condition used by Pure Type Systems \cite{Brn92}, to which we
add non-recursive untyped abbreviations, an infinite number of typed sorts, 
explicit type annotations, and some reduction schemes involving these
constructions. Remarkably we also replace the call-by-value
$\beta$-contraction scheme with its call-by-value version.
Then we show that the resulting typed $\lambda$-calculus, that we term $\LD$,
satisfies some important desirable properties such as the confluence of
reduction, the correctness of types, the uniqueness of types up to conversion,
the subject reduction of the type assignment, the strong normalization of
the typed terms and, as a corollary, the decidability of type inference
problem.

$\LD$ features the unification of terms and types, the immersion of
environments into terms, a ``compatible'' typing policy in which the dynamic
aspect of the type assignment is confined in the ``conversion rule'' and
finally a predicative abstraction.

The author conjectures that the expressive power of $\LD$ is that of $\LP$.

We see an application of this calculus as a formal specification language for
the type theories, like $\MTT$ \cite{MS05} or $\CTT$ \cite{NPS90,MLS84},
that require to be expressed in a predicative foundation.   
In this sense $\LD$ can be related both to PAL$^+$ \cite{Luo03} and to 
Martin-L\"of's theory of expressions \cite{NPS90}, that pursue the same aim and
use the type system of $\LR$ (i.e. they use arities).
Namely the author conjectures that $\LD$ includes both these theories.
In particular these calculi use $k$-uples of terms and $\LD$ can provide for 
this construction as well (see \appref{aggregates}).

The advantage of $\LD$ on these calculi is that the structural rules of
$\MTT$ and $\CTT$ can be justified by the rules of our calculus 
(see \appref{mtt}).

As an additional feature, the extension of $\LD$ termed $\CLD$ 
(\subsecref{exclusion}) comes with a full machine-checked specification of
its properties (see \subsecref{specification}).

In this section we will discuss some design features of $\CLD$ 
(\subsecref{blocks}) and we will summarize the open issues of the calculus
(\subsecref{todo}).

%% file: blocks.tex
\subsection{The Block Structure of $\CLD$}
\subseclabel{blocks}

$\CLD$ was carefully designed by the author on the basis of the criteria
discussed in \subsecref{outline}.
Another important design issue of this calculus is its \emph{block structure},
where by a \emph{block} we mean a subset of constructions and reduction rules
tightly connected to each other that we see as a unit (see \figref{blocks}{}).

\begin{figure}
\begin{center}

\begin{tabular}{|c|c|l|}
\hline
block id & main item & main item denomination \\ \hline
8 & $\VOID{x}{}$ & unconditioned exclusion \\ \hline
5 & $\ABST{x}{W}{}$ & abstraction over a complete type $W$ \\ \hline
1 & $\ABBR{x}{V}{}$ & unconditioned abbreviation of $V$ \\ \hline
0 & $\SORT{h}$ & sort of level $h$ \\ \hline
-1 & $\LREF{x}$ & variable occurrence \\ \hline
\end{tabular}

\end{center}
\caption{Block hierarchy in $\CLD$}
\figlabel{blocks}
\end{figure}

$\CLD$ has one block for each non-recursive construction and one for
each binder.

The author assigned a numeric identifier to each block just to suggest a
hierarchy in the block structure.
The type $W$ on which we abstract using $\ABST{x}{W}{}$ is \emph{complete}
because it represents a complete specification of the functional structure of
its inhabitants (see the comments on \figref{ty3}{abst}).
The abbreviation introduced by $\ABBR{x}{V}{}$ is \emph{unconditioned} because
it can always be unfolded by reduction. 

Generally speaking each binder has a \emph{domain} by which we mean the class
of the terms that can be substituted for the variable occurrences referring to
that binder.
Moreover a binder is here called \emph{conditioned} if it has an applicator
item associated to a specific reduction rule.
The applicator item always swaps with a binder of a different block by means
of a $\upsilon$ reduction step (see \figref{pr0}{} and \figref{pr0-void}{})
and the specific reduction rule always contracts the applicator-binder pair to
an unconditioned abbreviation.
An unconditioned binder is always eliminable by reduction when it is not an
environment entry.
if this domain is specified up to a non-trivial equivalence relation, its
inhabitants can be annotated with a preferred specification of this domain.
The annotator item can always be removed by reduction.

\begin{figure}
\begin{center}

\begin{tabular}{|l|l|l|l|l|l|}
\hline
item & domain & 
applicator with $\SR{\upsilon}$ &
reduction & $\SR{\zeta}$ & 
annotator with $\SR{\tau}$ \\ \hline

$\VOID{x}{}$ & $\css{T \st \top}$ &
no & no & yes & no \\ \hline

$\ABST{x}{V}{}$ & $\css{T \st \TyT{g}{E}{T}{V}}$ & 
$\APPL{V}{}$ & $\SR{\beta}$ & no &
$\CAST{W}{}$ \\ \hline 

$\ABBR{x}{V}{}$ & $\css{T \st T = V}$ & 
no & $\SR{\delta}$ & yes &
no \\ \hline
\end{tabular}

\end{center}
\caption{Detailed structure of the blocks about binding items}
\figlabel{binders}
\end{figure}

These considerations are summarized in \figref{binders}{} where the
$\lambda$-abstraction is considered in an environment $E$.
Notice that the abbreviation and the exclusion do not have an applicator
with a specific reduction because they are unconditioned. 

%% file: todo.tex
\subsection{Open Issues} 
\subseclabel{todo}

As already stressed along the paper, our presentation of $\LD$ leaves some
open issues that we want to reconsider in this subsection.

First of all, some technical aspects of the calculus need to be improved:
this includes taking a final decision on the shape of \defcref{arity}{} and of
\defcref{prc}{}. 

In particular we plan to reformulate the reduction predicates without the
explicit substitution
(\defcref{subst0}{}, \defcref{csubst0}{}, \defcref{fsubst0}{}).
and we want to reformulate the arity assignment without the level equality
(\defcref{leq}{}) that is undecidable in general.
We might also want to add the following type assignment rule:
\[
\Rule
{\TyT{g}{E}{T}{U} \quad \TyT{g}{E}{\APPL{V}{U}}{W}}{appl2}
{\TyT{g}{E}{\APPL{V}{T}}{\APPL{V}{U}}}
\]
with which we expect to type in $\LD$ all terms typable in $\LI$
(see \subsecref{outline}).

The items $\ABST{y}{D}{}$, $\ABBR{y}{F}{}$, $\APPL{F}{}$ and $\CAST{D}{}$
are not allowed at the moment (recall that $D$ and $F$ stand for
environments), but when $\LD$ will be extended by considering them as well,
a duality between terms and environments will arise (see \appref{duality}).

Secondly there are some conjectures that need to be proved formally.
In particular we are interested in understanding if the problem of type
inhabitation is decidable (this is an important property of $\LR$, see
\cite{Brn92}).

Thirdly we might want to extend $\CLD$ adding more blocks in the sense of
\subsecref{blocks}. 
Namely there are five constructions that can be of interest: 
declared constants (block 4), meta-variables (block -2), parameters,
(block 7), conditioned abbreviations (block 3) and abstractions over
incomplete types (block 6).

The first three constructions are taken from real implementations of typed
$\lambda$-calculus.
In particular we see the declaration of a constant as the unconditioned
version of the $\lambda$-abstraction, which we would like to denote with
$\DECL{x}{V}{}$ (where the $o$ can mean \emph{opaque} or can be an
\emph{omicron} chosen after $\acute{o} \nu o \mu \alpha $: Greek for ``name'').

Parameters appear in many logical frameworks \cite{KLN04,Luo03}.
Conditioned abbreviations are based on the binder $\DOOR{x}{V}{}$, on the
applicator $\CONN{V}{}$ and on the reduction rule
$\CONN{V}{\DOOR{x}{V}{T}} \SR{\beta c} \ABBR{x}{V}{T}$. 
They provide for possibly unexpandable abbreviations and mainly the applicator
$\CONN{V}{}$ does not carry any information into a $\beta c$-redex except for
its presence (since the term $V$ appears in the binder).
So we suspect that $\CONN{V}{}$ can be related to a \emph{connection}
of a Whole Adaptive System \cite{Sol05a} and we call $\CONN{V}{}$ a 
\emph{connessionistic} application item.%
\footnote{Describing the computational model of a Whole Adaptive System in
terms of a typed $\lambda$-calculus requires much more than conditioned
abbreviations: in particular we feel that anti-binders, in the sense of
\cite{HO03}, might play an important role for this task.}

Abstractions over incomplete types (i.e. types that do not specify the
functional structure of their inhabitants completely) are meant to simulate
the $\Pi$-abstractions of the $\lambda$-cube \cite{Brn92} and the author sees
fitting the $\Pi$ binder into the architecture of $\LD$ as a very challenging
task. In particular it would be interesting to relate this extension of
$\LD$ to COC since this calculus has been fully specified in \textsc{coq} 
\cite{Brr96} as well as $\LD$ itself, and the author sees the possibility of
certifying rigorously the mappings that may exist between these systems.

The novelty of $\LD$ extended with $\Pi$ would be that $\Pi$ could appear
at the level of terms and inside environments rather than only at the level of
types. 

In the perspective of relating this extension with a COC with universes, we
would also need a mechanism that makes $\SORT{h}$ a sub-sort of $\SORT{k}$
when $h < k$.


%% file: mtt.tex
\section{Justifying the Structural Fragment of $\MTT$ with $\LD$}
\applabel{mtt}

In the present appendix we show how the structural rules of Minimal Type
Theory ($\MTT$) \cite{MS05} can be justified trough the rules of $\LD$ and
we proceed in three steps. 
In \appref{expressions} we show that $\LD$ can be used as a theory of
expressions for $\MTT$.
In \appref{judgements} we show that $\LD$ type assignment and conversion
judgements can model $\MTT$ judgements.
In \appref{rules} we show that $\LD$ rules can model $\MTT$ structural rules.
In order to achieve this objective, we propose to remove $\eta$-conversion and
the so-called $\JCont{}$ judgement from $\MTT$, and to perform some changes to
the $\MTT$ rules called \emph{var} and \emph{prop-into-set}.

Our justification is based on a straight forward mapping of judgements, which
exploits uniformly dependent types on the $\LD$ side.
The underlying idea is to map the inhabitation judgements to the
type judgement $\TyT{}{}{}{}$ (at different levels of the type hierarchy) and
the equality judgements to the conversion judgement $\PcT{}{}{}$.
 
When referring to $\MTT$ we will use the notation of \cite{MS05}.

%% file: expressions.tex
\subsection{$\LD$ can serve as a Theory of Expressions for $\MTT$}
\applabel{expressions}

According to \cite{MS05} the theory of expressions underlying $\MTT$ is the
one, originally due to Martin-L\"of, underlying $\CTT$ \cite{NPS90} without
combinations and selections.
Moreover typed abstractions (\'a la Church) are used in place of untyped ones.

Therefore $\MTT$-expressions are based on variables, primitive constants,
defined constants, applications and typed abstractions.

Moreover every meaningful $\MTT$-expression has an \emph{arity}, which is a
type expression of the instance of $\LR$ with one type constant $0$.  

\indent\kern-0.2pt
Equality between $\MTT$-expressions is defined up to \emph{definitional
equality}: a rewriting mechanism that incorporates 
$\alpha\beta\eta$-conversion, and $\delta$-conversion 
(equality between the definiendum and the definiens of an abbreviation). 

In our proposal we leave $\eta$-conversion aside because we suspect that
this conversion is not strictly necessary in $\MTT$ and is used just as
syntactic sugar.
In any case $\eta$-conversion is available for $\lambda$-abstractions as
expected (see \thcref{prc_props}{pc3_eta}).

As a matter of fact $\LD$ can handle the mentioned ingredients as follows.

\textbf{Variables}, \textbf{defined constants}, \textbf{applications} and
\textbf{typed abstractions} are term constructions of the calculus
(see \defcref{TE}{}).
In particular we regard all definitions as $\delta$-entries of a
\emph{global environment} $E_y$ in which we close every term.

\textbf{Primitive constants} are regarded as references to $\lambda$-entries
(i.e. declarations) of the environment $E_y$.
So $E_y$ contains declarations and definitions.

\textbf{Types} can be substituted for arities.
Notice that arities exist in $\LD$ as well (see \defcref{arity}{})
and that typed terms have an arity (see \thcref{ty3_arity_props}{ty3_arity}).

Finally \textbf{definitional equality} is handled through environment-dependent
parallel conversion (see \defcref{prc}{}) that incorporates
$\alpha\beta\delta$-conversion.

%% file: judgements.tex
\subsection{$\LD$ Judgements can express $\MTT$ Judgements}
\applabel{judgements}

$\MTT$ features six main judgements that fall into two classes: declarations
and equalities.
Declarations state that an expression is a legal proposition, a legal data
type, or a legal element of a data type.
Equalities state that two legal propositions, data types, or elements of a data
type are semantically equal. 

Parametric expressions are allowed and each main judgement includes an
\emph{explicit environment} where the local parameters are declared. 

Other parameters, shared among all judgements of a given rule, are declared in
an \emph{implicit environment} extracted from the premises of that rule.

Summing up, a legal $\MTT$-expression requires three environments: the explicit
environment (provided by the judgement containing that expression), the
implicit environment (extracted from the premises of the rule containing that
judgement) and the global environment (for global declarations and
abbreviations).

A judgement stating that an explicit environment is legal, is also provided.

We can map these judgements to $\LD$-judgements in the way we explain below.

\vskip 2pt
\textbf{Sort hierarchy.} 
We need two sorts $\Prop$ and $\Set$ that we regard as aliases of $\SORT{0}$
and $\SORT{1}$ respectively (we can include these abbreviations in the global
environment $C_y$). We also set the sort hierarchy parameter 
(see \subsecref{native-defs}) to the function $\Gtz$ such that
$\Next{\Gtz}{h} \equiv h + 2$
This is the simplest choice ensuring that the positions of $\Set$ and $\Prop$
in the sort hierarchy graph (see \subsecref{arity-defs}) are disconnected.
In particular we observe that if $\Next{\Gz}{h} \equiv h + 1$ (as in
\subsecref{examples}) we derive directly from \figref{ty3}{sort}:
$\TyT{\Gz}{E}{\Set}{\Prop}$, which is against the intuition.

\vskip 2pt
\textbf{Environments.}
The explicit environment of an $\MTT$-judgement has the form: \\
$ \Gamma \equiv \JIn{}{x_1}{A_1}, \ldots, \JIn{}{x_n}{A_n} $
where $x_i$ is a variable and $A_i$ is an expression.

We can map each declaration of $\Gamma$ in a $\lambda$-entry, so
$\Gamma$ itself becomes the environment 
$C_x \equiv \ABST{x_1}{A_1}{} \ldots \ABST{x_n}{A_n}{\Set}$ of $\LD$.

The implicit environment of an $\MTT$-judgement does not need an explicit
mapping since we can exploit the implicit environment of the corresponding
judgement of $\LD$
(at least as long as we are dealing just with the structural rules of $\MTT$).

\vskip 2pt
\textbf{Declarations:} 
$\JProp{\Gamma}{A}$,
$\JSet{\Gamma}{A}$,
$\JIn{\Gamma}{a}{A}$,
$\JCont{\Gamma}$.

A declaration judgement is mapped to a type assignment judgement
(see \defcref{ty3}{}). Namely we map 
$\JProp{\Gamma}{A}$ to $\TyT{\Gtz}{\PUSH{C_y}{C_x}}{A}{\Prop}$, we map 
$\JSet{\Gamma}{A}$ to $\TyT{\Gtz}{\PUSH{C_y}{C_x}}{A}{\Set}$ and we map 
$\JIn{\Gamma}{a}{A}$ to $\TyT{\Gtz}{\PUSH{C_y}{C_x}}{a}{A}$ in the implicit
environment $\TyT{\Gtz}{\PUSH{C_y}{C_x}}{A}{\Set}$.
Here $\PUSH{C_y}{C_x}$ refers to the concatenation of $C_y$ and $C_x$.
Notice that the type assignment is invariant for conversion
(modelling definitional equality) as stated by 
\figref{ty3}{\mathrm{conv}} and \thcref{ty3_sred}{ty3_sconv}.

Coming to the legal explicit environment judgement $\JCont{\Gamma}$, the experience
of the author with $\LD$ shows that such a judgement is useless (as it does
not guarantee additional meta-theoretical properties) and heavy (as it
introduces a mutual dependence between itself and $\JSet{\Gamma}{A}$ at the
meta-theory level). 
The point is that an unreferenced parameter does not need a legal declaration
unless it is the formal argument of a function. 
So we propose not to map $\JCont{\Gamma}$ and to change the related rules
(see \appref{rules}).
In any case legal environments are supported in the calculus $\CLD$
(\subsecref{legal}) if they are needed for some reason.

\vskip 2pt
\textbf{Equalities:}
$\JPEq{\Gamma}{A_1}{A_2}$,
$\JSEq{\Gamma}{A_1}{A_2}$, 
$\JEq{\Gamma}{a_1}{a_2}{A}$.

An equality judgement is mapped to an environment-dependent conversion
judgement (see \defcref{prc}{}). Namely, we map 
$\JKEq{\Gamma}{A_1}{A_2}{S}$ to $\PcT{\PUSH{C_y}{C_x}}{A_1}{A_2}$
in the implicit environment 
$\TyT{\Gtz}{\PUSH{C_y}{C_x}}{A_1}{S}$ and
$\TyT{\Gtz}{\PUSH{C_y}{C_x}}{A_2}{S}$
where $S$ is either $\Prop$ or $\Set$, and we map 
$\JEq{\Gamma}{a_1}{a_2}{A}$ to $\PcT{\PUSH{C_y}{C_x}}{a_1}{a_2}$
in the implicit environment
$\TyT{\Gtz}{\PUSH{C_y}{C_x}}{a_1}{A}$, 
$\TyT{\Gtz}{\PUSH{C_y}{C_x}}{a_2}{A}$
and $\TyT{\Gtz}{\PUSH{C_y}{C_x}}{A}{\Set}$.

Notice that the conversion judgement is invariant for conversion itself
(modelling definitional equality) because the conversion is an equivalence
relation. 

%% file: rules.tex
\subsection{$\LD$ Rules can express $\MTT$ Structural Rules}
\applabel{rules}

Our proposal for the structural rules of $\MTT$ is shown in \figref{mtt}{}.

\begin{figure}
\begin{center}

\Rule{\JProp{\Gamma}{A}}{\mathrm{ps}}{\JSet{\Gamma}{\Proof{A}}} \quad
\Rule
 {\JSet{\Gamma}{A}}{\mathrm{var}}
 {\JIn{\Gamma,\JIn{}{x}{A},\Delta}{\MLRef{x}}{A}} \\
\Rule
 {\JIn{\Gamma}{a}{A_1} \quad \JSEq{\Gamma}{A_1}{A_2}}
 {\mathrm{seteq}}{\JIn{\Gamma}{a}{A_2}} \\
\Rule{\JSet{\Gamma}{A}}{\mathrm{r}}{\JSEq{\Gamma}{A}{A}} \quad
\Rule{\JSEq{\Gamma}{A_1}{A_2}}{\mathrm{s}}{\JSEq{\Gamma}{A_2}{A_1}} \quad
\Rule
 {\JSEq{\Gamma}{A_1}{A} \quad \JSEq{\Gamma}{A}{A_2}}
 {\mathrm{t}}{\JSEq{\Gamma}{A_1}{A_2}} \\
\Rule{\JProp{\Gamma}{A}}{\mathrm{r}}{\JPEq{\Gamma}{A}{A}} \quad
\Rule{\JPEq{\Gamma}{A_1}{A_2}}{\mathrm{s}}{\JPEq{\Gamma}{A_2}{A_1}} \quad
\Rule
 {\JPEq{\Gamma}{A_1}{A} \quad \JPEq{\Gamma}{A}{A_2}}
 {\mathrm{t}}{\JPEq{\Gamma}{A_1}{A_2}} \\
\Rule{\JIn{\Gamma}{a}{A}}{\mathrm{r}}{\JEq{\Gamma}{a}{a}{A}} \quad
\Rule{\JEq{\Gamma}{a_1}{a_2}{A}}{\mathrm{s}}{\JEq{\Gamma}{a_2}{a_1}{A}} \quad
\Rule
 {\JEq{\Gamma}{a_1}{a}{A} \quad \JEq{\Gamma}{a}{a_2}{A}}
 {\mathrm{t}}{\JEq{\Gamma}{a_1}{a_2}{A}} \\
\Rule
 {\JSet{\Gamma}{A} \quad \JSet{\Gamma,\JIn{}{x}{A}}{B}}{\mathrm{i}}
 {\JKind{\Gamma}{\MAbst{x}{A} B}{\MAbst{x}{A} \Set}} \quad
\Rule
 {\JSet{\Gamma}{A} \quad \JSEq{\Gamma,\JIn{}{x}{A}}{B_1}{B_2}}{\mathrm{i}}
 {\JKEq{\Gamma}{\MAbst{x}{A} B_1}{\MAbst{x}{A} B_2}{\MAbst{x}{A} \Set}} \\
\Rule
 {\JSet{\Gamma}{A} \quad \JProp{\Gamma,\JIn{}{x}{A}}{B}}{\mathrm{i}}
 {\JKind{\Gamma}{\MAbst{x}{A} B}{\MAbst{x}{A} \Prop}} \quad
\Rule
 {\JSet{\Gamma}{A} \quad \JPEq{\Gamma,\JIn{}{x}{A}}{B_1}{B_2}}{\mathrm{i}}
 {\JKEq{\Gamma}{\MAbst{x}{A} B_1}{\MAbst{x}{A} B_2}{\MAbst{x}{A} \Prop}} \\
\Rule
 {\JSet{\Gamma}{A} \quad 
  \JIn{\Gamma,\JIn{}{x}{A}}{b}{B}}{\mathrm{i}}
 {\JKIn{\Gamma}{\MAbst{x}{A} b}{\MAbst{x}{A} B}{\MAbst{x}{A} \Set}} \quad
\Rule
 {\JSet{\Gamma}{A} \quad \JEq{\Gamma,\JIn{}{x}{A}}{b_1}{b_2}{B}}
 {\mathrm{i}}
 {\JKIEq{\Gamma}{\MAbst{x}{A} b_1}{\MAbst{x}{A} b_2}{\MAbst{x}{A} B}
        {\MAbst{x}{A} \Set}} \\
\Rule
 {\JIn{\Gamma}{a}{A} \quad\JKind{\Gamma}{B}{\MAbst{x}{A} \Set}}
 {\mathrm{e}}{\JSet{\Gamma}{\MAppl{B}{a}}} \quad
\Rule
 {\JIn{\Gamma}{a}{A} \quad
  \JKEq{\Gamma}{B_1}{B_2}{\MAbst{x}{A} \Set}}
 {\mathrm{e}}{\JSEq{\Gamma}{\MAppl{B_1}{a}}{\MAppl{B_2}{a}}} \\
\Rule
 {\JIn{\Gamma}{a}{A} \quad \JKind{\Gamma}{B}{\MAbst{x}{A} \Prop}}
 {\mathrm{e}}{\JProp{\Gamma}{\MAppl{B}{a}}} \quad
\Rule
 {\JIn{\Gamma}{a}{A} \quad
  \JKEq{\Gamma}{B_1}{B_2}{\MAbst{x}{A} \Prop}}
 {\mathrm{e}}{\JPEq{\Gamma}{\MAppl{B_1}{a}}{\MAppl{B_2}{a}}} \\
\Rule
 {\JIn{\Gamma}{a}{A} \quad
  \JKIn{\Gamma}{b}{\MAbst{x}{A} B}{\MAbst{x}{A} \Set}}
 {\mathrm{e}}{\JIn{\Gamma}{\MAppl{b}{a}}{\MAbst{x}{A}\MAppl{B}{a}}} \\
\Rule
 {\JIn{\Gamma}{a}{A} \quad
  \JKIEq{\Gamma}{b_1}{b_2}{\MAbst{x}{A} B}{\MAbst{x}{A} \Set}}
 {\mathrm{e}}{\JEq{\Gamma}{\MAppl{b_1}{a}}{\MAppl{b_2}{a}}{\MAbst{x}{A}\MAppl{B}{a}}}
	
\end{center}
\caption{Our proposal for the structural rules of $\MTT$}
\figlabel{mtt}
\end{figure}

\textbf{the \emph{prop-into-set} rule} can not be modelled, as it is, by $\LD$
because $\LD$ does not feature subtyping. Therefore our proposal is to make
the coercion from $\Prop$ to $\Set$ explicit. Namely we declare a primitive
constant $\Proof{}$ of type $\ABST{x}{\Prop}{\Set}$ in the global environment
$C_y$ and we set \figref{mtt}{\mathrm{ps}} modelled by \figref{ty3}{appl}.
This solution is well known in the literature (see \cite{CH88,SPAb1,SPAa2}).

\textbf{The \emph{var} rule.}
Our proposal for this rule is \figref{mtt}{\mathrm{var}}
modelled by \figref{ty3}{\mathrm{decl}}.
The implicit environment is respected because of
\thcref{ty3_props}{ty3_lift}.

\textbf{The \emph{seteq} rule.}
This rule is \figref{mtt}{\mathrm{seteq}}
modelled by \figref{ty3}{\mathrm{conv}} whose first premise is taken from the
implicit environment.

\textbf{The equivalence rules} of the equality judgements are justified by
the fact that the environment-dependent conversion is an equivalence relation.

The complete list is in \figref{mtt}{} (labels: $\mathrm{r}$, $\mathrm{s}$,
$\mathrm{t}$).

\textbf{The derivable rules.}
Notice that \cite{NPS90} suggests some additional structural rules (like a
second $\mathrm{seteq}$ rule and some substitution rules) that are not
included in $\MTT$ because they are derivable.
In the $\LD$ perspective we derive these rules from
\thcref{prc_props}{pc3_thin_dx_appl}, \thcref{prc_props}{pc3_head_1},
\thcref{ty3_props}{ty3_subst0} and \thcref{ty3_props}{csubt_ty3}.

\textbf{The rules on classes.}
If we regard $\Prop$ and $\Set$ as primitive constants rather than judgement
keywords, we can build expressions like
$\MAbst{x_1}{e_1} \ldots \MAbst{x_n}{e_n} \Set$
or $\MAbst{x_1}{e_1} \ldots \MAbst{x_n}{e_n} \Prop$
(called \emph{types} in $\MTT$ or \emph{categories} in $\CTT$ \cite{MLS84}).
With these ``classes'' we can form the following judgements:
\[\begin{tabular}{ll}
$\JKind{\Gamma}{B}{\MAbst{x}{A} \Set}$ &
$\JKEq{\Gamma}{B_1}{B_2}{\MAbst{x}{A} \Set}$ \\
$\JKind{\Gamma}{B}{\MAbst{x}{A} \Prop}$ &
$\JKEq{\Gamma}{B_1}{B_2}{\MAbst{x}{A} \Prop}$ \\
$\JKIn{\Gamma}{b}{B}{\MAbst{x}{A} \Set}$ &
$\JKIEq{\Gamma}{b_1}{b_2}{B}{\MAbst{x}{A} \Set}$
\end{tabular}\]
that we explain with the rules modelled by
\figref{ty3}{abst} and \thcref{prc_props}{pc3_head_2}.
These rules are shown in \figref{mtt}{} with the label: $\mathrm{i}$.
The elimination rules, modelled by \figref{ty3}{appl} and
\thcref{prc_props}{pc3_thin_dx_appl}, are shown in \figref{mtt}{} with the
label: $\mathrm{e}$.

%% file: duality.tex
\section{Towards a Duality between Terms and Environments}
\applabel{duality}

The present appendix contains some hints on how the author plans to complete
$\LD$ by adding the items $\ABST{y}{D}{}$, $\ABBR{y}{F}{}$, $\APPL{F}{}$ and
$\CAST{D}{}$ both in the terms and in the environments.
In principle the need for these items was evident from the very start but they
were not included in \cite{Gui05} because of the technical problems they
seemed to give. 
In particular the author did not see the importance of the iterated static
type assignment as a way to map $\TRM$ into $\ENV$ (\subsecref{static-defs})
until the properties of $\LD$ were made clear 
(especially \thcref{sty_props}{sty1_env}, \thcref{sty_props}{ty3_sty0}
and \thcref{ty3_arity_props}{ty3_arity}).
We would like to stress that the contents of this appendix are just a proposal
for future research on $\LD$ and have not been certified yet.

In \appref{complete} we introduce these new items,
In \appref{aggregates} we propose the new term construction $\PROJ{F}{T}$ as
an application, 
in \appref{polarity} we propose to merge $\TRM$ and $\ENV$ in a single data
type to avoid the replication of dual definitions and theorems in the
perspective of certifying the properties of complete $\LD$.

%% file: complete.tex
\subsection{Complete $\LD$: Dualizing Terms and Environments}
\applabel{complete}

According to \defcref{TE}{} the argument of the abstractors, abbreviators,
applicators and type annotators is a term.
Nevertheless an environment can be allowed as well.

\begin{definition}[complete syntax of terms and environments]\hfil
\objlabel{CTE}

The complete versions of $\TRM$ and $\ENV$ are defined by extending
\defcref{TE}{} as follows:
\[\TRM \equiv \TRM \| 
              \ABST{\WAR}{\ENV}{\TRM} \| \ABBR{\WAR}{\ENV}{\TRM} \| 
	      \APPL{\ENV}{\TRM} \| \CAST{\ENV}{\TRM} 
\]
\[\ENV \equiv \ENV \| \LREF{\WAR} \|
              \ABST{\WAR}{\ENV}{\ENV} \| \ABBR{\WAR}{\ENV}{\ENV} \| 
	      \APPL{\ENV}{\ENV} \| \CAST{\ENV}{\ENV}
\]
where $\WAR$ is a set of names for variables denoting environments.

We call a recursive construction positive when its arguments belong to the
same type and negative otherwise.
We call this attribute the polarity of the construction.

\end{definition}

Notice that the calculi of the $\LM$ family use two different sets of
variables as well.

Once defined in this way, $\TRM$ and $\ENV$ are isomorphic through the
polarity preserving transformations $\E{} \of \TRM \to \ENV$ and 
$\T{} \of \ENV \to \TRM$ defined below.

\begin{definition}[the transformations $\E{}$ and $\T{}$]\hfil

The transformations
$\E{} \of \TRM \to \ENV$ and $\T{} \of \ENV \to \TRM$
work as follows:

\begin{enumerate}\EnumStyle

\item
$ \E{\SORT{h}} = \SORT{h} $ and $ \T{\SORT{h}} = \SORT{h} $;

\item
$ \E{\LREF{x}} = \LREF{y} $ and $ \T{\LREF{y}} = \LREF{x} $
(here we assume that $\VAR$ and $\WAR$ are isomorphic);

\item
$ \E{\ABST{x}{W}{T}} = \ABST{y}{\E{W}}{\E{T}} $ and
$ \T{\ABST{y}{D}{E}} = \ABST{x}{\T{D}}{\T{E}} $;

\item
$ \E{\ABST{y}{D}{T}} = \ABST{x}{\T{D}}{\E{T}} $ and
$ \T{\ABST{x}{W}{E}} = \ABST{y}{\E{W}}{\T{E}} $;

\item
$ \E{\ABBR{x}{V}{T}} = \ABBR{y}{\E{V}}{\E{T}} $ and
$ \T{\ABBR{y}{F}{E}} = \ABBR{x}{\T{F}}{\T{E}} $;

\item
$ \E{\ABBR{y}{F}{T}} = \ABBR{x}{\T{F}}{\E{T}} $ and
$ \T{\ABBR{x}{V}{E}} = \ABBR{y}{\E{V}}{\T{E}} $;

\item
$ \E{\APPL{V}{T}} = \APPL{\E{V}}{\E{T}} $ and
$ \T{\APPL{F}{E}} = \APPL{\T{F}}{\T{E}} $;

\item
$ \E{\APPL{F}{T}} = \APPL{\T{F}}{\E{T}} $ and
$ \T{\APPL{V}{E}} = \APPL{\E{V}}{\T{E}} $;

\item
$ \E{\CAST{W}{T}} = \CAST{\E{W}}{\E{T}} $ and
$ \T{\CAST{D}{E}} = \CAST{\T{D}}{\T{E}} $;

\item
$ \E{\CAST{D}{T}} = \CAST{\T{D}}{\E{T}} $ and
$ \T{\CAST{W}{E}} = \CAST{\E{W}}{\T{E}} $.

\end{enumerate}

\end{definition}

\defcref{CTE}{} opens some issues: we discuss the most relevant below.

\textbf{Focalized terms.}
When a term reference $\LREF{x}$ points to an abstractor $\ABST{x}{W}{}$ in an
environment $E$ it may be the case that the rightmost item of $E$ is not a
sort. In that event we must consider its iterated static type 
(see \thcref{sty_props}{sty1_env}).
More precisely if $E$ is $\PUSH{C}{\LREF{y}}$ and if $\LREF{y}$ points to
$\ABST{y}{D}{}$ or to $\ABBR{y}{F}{}$, we recursively resolve $\LREF{x}$ in
the environments $\PUSH{C}{D}$ or $\PUSH{C}{F}$ respectively (this is much
like considering the iterated static type of $E$ except for the rightmost sort
item that is irrelevant when searching for binders).
This solution may look strange at a first glance but consider 
$E = \ABST{y}{D}{\LREF{y}}$: this is the empty environment whose ``hole'' is
$y$ in the sense of \cite{CH00}.
Normally the references to the empty environment are not legal but in our case
the ``hole'' is typed explicitly so we can foresee its contents by inspecting
its type. This means that for $D = \ABST{x}{W}{\SORT{\Next{g}{n}}}$ the
focalized term $(\ABST{y}{D}{\LREF{y}}, \LREF{x})$ is legal and the term
reference $\LREF{x}$ points to $\ABST{x}{W}{}$. Furthermore that reference
continues to point to the same binder when $E$ is instantiated and reduced:

\begin{enumerate}

\item
Legal instantiation with $F = \ABST{x}{W}{\SORT{n}}$: 
$(\APPL{F}{\ABST{y}{D}}{\LREF{y}}, \LREF{x})$.

\item
$\beta$-contraction: 
$(\ABBR{y}{F}{\LREF{y}}, \LREF{x})$.

\item
$\delta$-expansion: 
$(\ABBR{y}{F}{F}, \LREF{x})$.

\item
$\zeta$-contraction: 
$(F, \LREF{x})$.

\end{enumerate}

As we see, everything works fine because the item $\ABST{x}{W}{}$ must appear
in $F$ as well as in $D$ in order for the instantiation to be legal (i.e. well
typed).

If the term reference $\LREF{x}$ points to an abbreviator $\ABBR{x}{V}{}$, we
do the same thing.

\textbf{Pushing.}
When moving an abstractor $\ABST{x}{W}{}$ from a term to an environment, as we
might need to do when the term and the environment themselves are the
components of a focalized term, we must make sure that the references to
$\ABST{x}{W}{}$ are preserved.
So, when the environment has the form $\PUSH{C}{\LREF{y}}$ where $\LREF{y}$
points to $\ABST{y}{D}{}$ or to $\ABBR{y}{F}{}$, we must move $\ABST{x}{W}{}$
recursively into $D$ or $F$ respectively.
In the first case this amounts to updating the explicit type $D$ of the
environment ``hole'' in a way that makes it possible to fill that ``hole''
through a legal instantiation.

As before, when we move an abbreviator $\ABBR{x}{V}{}$, we do the same thing.

\textbf{Reduction.}
The $\beta$-redexes are $\APPL{V}{\ABST{x}{W}{}}$
(from \subsecref{reduction-defs}) and symmetrically $\APPL{F}{\ABST{y}{D}{}}$.
The abbreviations $\ABBR{x}{V}{E}$ do not $\zeta$-reduce (from \cite{Gui06a})
and symmetrically the abbreviations $\ABBR{y}{F}{T}$ do not $\zeta$-reduce
either.

%% file: aggregates.tex
\subsection{Environments as Aggregates}
\applabel{aggregates}

Formally the $k$-uple $(V_{k-1}, \ldots, V_0)$ at position $\NODE{0}{h}$ in
the type hierarchy is denoted by the environment
$E = \ABBR{x_{k-1}}{V_{k-1}}{} \ldots \ABBR{x_0}{V_0}{\SORT{h}}$.

More generally the binders $\ABST{x}{W}{}$ and $\ABBR{x}{V}{}$ of an
environment $E$ (as well as the binders $\ABST{y}{D}{}$ and
$\ABBR{y}{F}{}$ of a term $T$) can be seen as the fields of an aggregate
structure.
These fields can be definitions (denoted by the $\ABBR{x}{V}{}$ items) or
declarations (denoted by the $\ABST{x}{W}{}$ items) and can be dependent. 
In order to be effective, aggregates need a projection mechanism that allows
to reed their fields.
To this aim we propose the item $\PROJ{F}{}$ that we call \emph{projector} and
the term constructions $\PROJ{F}{T}$ that we call \emph{projection}.
Considering the previous $k$-uple $E$, the basic idea is that
$\PROJ{E}{\LREF{x_i}}$ must reduce to $V_i$, so we set the following
sequential reduction rule.
\[\hbox{
If $F \mathrel\vdash T_1 \mathrel\rightarrow T_2$ and if $T_2$ does not
refer to $F$ then $\PROJ{F}{T_1} \SR{\pi} T_2$
}\]

Notice that $\PROJ{F}{T}$ might be related to the \emph{with} instruction of
the \textsc{pascal} programming language \cite{JW81} and might look like:
\emph{with} $F$ \emph{do} $T$.

Following the ``environments as aggregates'' interpretation, we might expect
to type $E$ with
$C_1 = \ABST{x_{k-1}}{W_{k-1}}{} \ldots \ABST{x_0}{W_0}{\SORT{\Next{g}{h}}}$ 
where each $W_i$ is the type of $V_i$.
Nevertheless the type of $E$ as a term is
$C_2 = \ABBR{x_{k-1}}{V_{k-1}}{} \ldots \ABBR{x_0}{V_0}{\SORT{\Next{g}{h}}}$
according to \defcref{ty3}{} but notice that $\Csubt{g}{C_1}{C_2}$
(this is the domain-based preorder of \subsecref{preorders}).
This consideration shows that it could make sense to investigate the
extension of $\LD$ with a subtyping relation based on $\Csubt{g}{}{}$.

%% file: polarity.tex
\subsection{Unified $\LD$: Introducing Polarized Terms}
\applabel{polarity}

In this subsection we propose the notion of a \emph{polarized term}: an
expression capable of representing both a term and an environment (in the
sense of \defcref{CTE}{}) in a way that turns the transformations $\E{}$ and
$\T{}$ into the identity functions.

The basic idea consists in decorating the recursive term constructions
with the information on their polarity represented as a boolean value.

Let us denote the data type of the boolean values with 
$\BOOLE \equiv \css{\BOT, \TOP}$
and let us assume that $\TOP$ (positive polarity) represents $\top$,
then a polarized term is as follows.

\begin{definition}[syntax of polarized terms]\hfil
\objlabel{P}

The set of polarized terms is defined as follows:
\[\POL \equiv \SORT{\NAT} \| \LREF{\VAR} \| 
              \BOOLE\ABST{\VAR}{\POL}{\POL} \| \BOOLE\ABBR{\VAR}{\POL}{\POL} \|
              \BOOLE\APPL{\POL}{\POL} \| \BOOLE\CAST{\POL}{\POL} 
\]

\end{definition}

\defcref{P}{} opens the issue of deciding whether a $Q \in \POL$ can be mapped
back to a $ V \in \TRM $ or to an $ F \in \ENV $.
Clearly the fact that the transformations $\E{}$ and $\T{}$ are mapped to
the identity functions on $\POL$ says that this information, which we call
the \emph{absolute polarity} of $Q$, is not recoverable.
What we can recover is the \emph{relative polarity} of $Q$ with respect to 
a superterm $P$ of $Q$ 
This is to say that we can know if $P$ and $Q$ represent two elements of
the same type or not.

\begin{definition}[relative polarity assignment]\hfil
\objlabel{Polarity}

The partial function $\Polarity{P}{Q}{}$, that returns $\TOP$ if the terms
$P$ and $Q$ have the same absolute polarity, is defined by the clauses shown
below, where $\Beq{}{}$ denotes the boolean coimplication
(i.e. the negated exclusive disjunction).

\begin{enumerate}

\item (refl)
$\Polarity{P}{P}{\TOP}$;

\item (trans)
if $\Polarity{P_1}{P}{b_1}$ and $\Polarity{P}{P_2}{b_2}$ then \\
$\Polarity{P_1}{P_2}{\Beq{b_1}{b_2}}$;

\item (fst)
$\Polarity{b\ABST{z}{Q}{P}}{P}{\TOP}$; 
$\Polarity{b\ABST{z}{Q}{P}}{P}{\TOP}$; \\
$\Polarity{b\APPL{Q}{P}}{P}{\TOP}$;
$\Polarity{b\CAST{Q}{P}}{P}{\TOP}$;

\item (snd)
$\Polarity{b\ABST{z}{Q}{P}}{Q}{b}$;
$\Polarity{b\ABBR{z}{Q}{P}}{Q}{b}$; \\
$\Polarity{b\APPL{Q}{P}}{Q}{b}$;
$\Polarity{b\CAST{Q}{P}}{Q}{b}$.

\end{enumerate}

\end{definition}

We conjecture that the knowledge of relative polarity is enough to treat the
version of $\LD$ based on polarized terms.
We call this calculus \emph{unified $\LD$} or $\ULD$.

As an example let us consider the restrictions on reduction mentioned in
\appref{complete}.
The unified $\beta$-redex takes the form $b\APPL{Q_1}{b\ABST{z}{Q_2}{}}$, while
$\zeta$-reduction is allowed on the items $\TOP\ABBR{z}{Q}{}$ and not allowed
on the items $\BOT\ABBR{z}{Q}{}$.

%% file: progress.tex
\section{A Note on the Current State of the Formal Specification}
\applabel{progress}
\lifttrue

In this appendix we discuss the current state of the definitions that formally
specify $\CLD$ in the $\CIC$ \cite{Gui05} in terms of modifications with
respect to their initial state \cite{Gui06a}.

Firstly we set up a mechanism to avoid the need of exchanging the environment
binders in the proof of \thcref{sn3_props}{sc3_arity_csubc}.
In particular we defined an extension of the \texttt{lift} function and an
extension of the \texttt{drop} function \cite{Gui06a} that apply a finite
number of relocations to a term.
The ``relocation parameters'' (i.e. the arguments $h$ and $i$ of the
\texttt{lift} function) are contained in a list of pairs $(h, i)$.
Here $\overline s$ will always denote a variable for such a list.

These definitions are given in \defcref{lift1}{} and \defcref{drop1}{} below.

\begin{definition}[the multiple relocation function]\hfil
\objlabel{lift1}
\[\left\{\begin{tabular}{lll}
$\LiftF{\PNil}{T}$&
$\equiv$&
$\Alone{T}$\\
$\LiftF{\PCons{h}{i}{\overline{s}}}{T}$&
$\equiv$&
$\Lift{h}{i}{\LiftF{\overline{s}}{T}}$\\
\end{tabular}\right.\]
\end{definition}

\begin{definition}[axioms for multiple dropping]\hfil
\objlabel{drop1}

\begin{enumerate}\EnumStyle

\item (non recursive case)
\objlabel{drop1_nil}

$\DropF{\PNil}{C}{C}$.

\item (recursive case)
\objlabel{drop1_cons}

If $\Drop{h}{i}{C_1}{C_2}$ and
$\DropF{\overline{s}}{C_2}{C_3}$ then
$\DropF{\PCons{h}{i}{\overline{s}}}{C_1}{C_3}$.

\end{enumerate}

\end{definition}

With these functions we were able to rephrase \defcref{sc3}{} as follows:

\begin{definition}[the strong reducibility predicate]\hfil
\objlabel{sc3-real}
\begin{footnotesize}
\[\left\{\begin{tabular}{@{\kern-2pt}lll}
$\ScT{g}{\ASort{k}{h}}{C}{T}$&
iff&
$\Arity{g}{C}{T}{\ASort{k}{h}}$ and
$\SnT{C}{T}$\\
$\ScT{g}{\AHead{L_1}{L_2}}{C}{T}$&
iff&
$\Arity{g}{C}{T}{\AHead{L_1}{L_2}}$ and
for each $D$, $W$, $\overline{s}$,\\
&
&
$\ScT{g}{L_1}{D}{W}$ and $\DropF{\overline{s}}{D}{C}$ imply
$\ScT{g}{L_2}{D}{\APPL{W}{\LiftF{\overline{s}}{T}}}$\\
\end{tabular}\right.\]
\end{footnotesize}
\end{definition}

The other definitions not included in \cite{Gui06a} were formalized
substantially as they appear in the previous sections, and we omit them here.

Remarkably we made some corrections to the preorders on environments
(\defcref{csubt}{}, \defcref{csuba}{}, \defcref{csubc}{}) in order to prove
\thcref{ty3_arity_props}{csubt_csuba}.

Notice that relocations (i.e. applications of the \texttt{lift} function) were
added where necessary both in the definitions and the theorems because in
\cite{Gui05}, variables are referenced by position and not by name as in the
present paper.

Secondly we took a final decision about the notation of the cast item, for
which we now use $\CAST{V}{}$ instead of $\PROJ{V}{}$ (see \defcref{TE}{},
\defcref{CTE}{} and \defcref{P}{}).
We also changed the native type assignment rule \figref{ty3}{cast} because
the former version applies an $\tau$-reduction at the level of types in
contrast with the general policy stated in \subsecref{outline}.
\thcref{ty3_gen}{ty3_gen_cast} is changed accordingly. 

Thirdly we took a final decision on the domain of the exclusion binder and we
rearranged the overall architecture of the calculus, also inserting the block
for declared constants (see \subsecref{blocks} and \subsecref{todo}). 

At the same moment we took a final decision on the name of the extension of
$\LD$ with the unconditioned exclusion binder, which is now $\CLD$ instead of
$\LD\chi$.

Finally we used $\SR{\tau}$ here in place of $\SR{\epsilon}$ for the reduction
step that removes explicit type casts to avoid a clash with other reduction 
steps named $\epsilon$ appearing in the literature (see for instance the
calculus $\lambda\epsilon$ in \cite{SU06}).

Currently (May 2008), the \emph{Basic} module of the certified specification
\cite{Gui05} consists of 525 kilobytes of \textsc{coq} vernacular describing
85 definitions and 683 theorems.
The \emph{Ground} module, that extends the standard library of \textsc{coq},
consists of 34 kilobytes of vernacular describing 28 definitions and 50
theorems.
From the standard library of \textsc{coq} we borrow 18 definitions and 69
theorems.

%% file: proofs.tex
\section{Pointers to the Certified Proofs}
\applabel{proofs}

As we mentioned in \subsecref{specification} the certified proofs of all
results stated in this paper are available as resources of the \HELM\
(\textsc{helm}) and their representation in natural language can be obtained
through the \textsc{helm} rendering software.
Each proof is identified by a path that we list below.

We provide two methods to obtain the representation of a proof:

\begin{itemize}

\item
The \textbf{dynamic representation} is generated on the fly by the
\textsc{helm} rendering software, which is very slow when big proofs are
rendered (such as \thcref{prc_props}{pr0_subst0}).

Visit the \textsc{helm} on-line library at \HelmBrowse, follow the path
\basepath\ and then the path of the proof.

You can not reach a proof by concatenating these paths in a single \url{http}
address.

\item
The \textbf{static representation} has been already generated so it displays
faster. 

Visit the $\LD$ web site at \LDStatic, follow the path \basepath\ and then the
path of the proof, that in this case has \url{.html} appended at the end.
You can also reach the proof by concatenating these three paths in a single
\url{http} address.

\end{itemize}

The proofs are displayed correctly only selecting a font with \textsc{Unicode}
support.

The following paths are parts of Uniform Resource Identifiers (\textsc{uri})
\cite{URI} so we can not guarantee their persistence.

\begin{enumerate}

\pointer{arity_props}{node_inh}{arity/props/node\_inh.con}
\pointer{arity_props}{arity_mono}{arity/props/arity\_mono.con}
\pointer{arity_props}{arity_fsubst0}{arity/subst0/arity\_fsubst0.con}
\pointer{arity_props}{csuba_arity}{csuba/arity/csuba\_arity.con}

\pointer{arity_sred}{arity_sred_wcpr0_pr0}{arity/pr3/arity\_sred\_wcpr0\_pr0.con}
\pointer{arity_sred}{arity_sred_pr3}{arity/pr3/arity\_sred\_pr3.con}

\pointer{prc_props}{pr0_subst0}{pr0/props/pr0\_subst0.con}
\pointer{prc_props}{pr0_confluence}{pr0/pr0/pr0\_confluence.con}
\pointer{prc_props}{pr3_confluence}{pr3/pr3/pr3\_confluence.con}
\pointer{prc_props}{pc3_thin_dx_appl}{pc3/props/pc3\_thin\_dx.con}
\pointer{prc_props}{pc3_head_1}{pc3/props/pc3\_head\_1.con}
\pointer{prc_props}{pc3_head_2}{pc3/props/pc3\_head\_2.con}
\pointer{prc_props}{pc3_gen_abst}{pc3/fwd/pc3\_gen\_abst.con}
\pointer{prc_props}{pc3_eta}{pc3/props/pc3\_eta.con}

\pointer{arity_nf2_inv_all}{}{nf2/arity/arity\_nf2\_inv\_all.con}

\pointer{csubc_props}{csubc_csuba}{csubc/csuba/csubc\_csuba.com}
\pointer{csubc_props}{csubc_arity_conf}{csubc/arity/csubc\_arity\_conf.com}

\pointer{sn3_props}{sn3_nf2}{sn3/nf2/sn3\_nf2.con}
\pointer{sn3_props}{sc3_cast}{sc3/props/sc3\_cast.con}
\pointer{sn3_props}{sc3_abbr}{sc3/props/sc3\_abbr.con}
\pointer{sn3_props}{sc3_abst}{sc3/props/sc3\_abst.con}
\pointer{sn3_props}{sc3_sn3}{sc3/props/sc3\_sn3.con}
\pointer{sn3_props}{sc3_bind}{sc3/props/sc3\_bind.con}
\pointer{sn3_props}{sc3_appl}{sc3/props/sc3\_appl.con}
\pointer{sn3_props}{sc3_arity_csubc}{sc3/arity/sc3\_arity\_csubc.con}
\pointer{sn3_props}{sc3_arity}{sc3/arity/sc3\_arity.con}

\pointer{ty3_gen}{ty3_gen_sort}{ty3/fwd/ty3\_gen\_sort.con}
\pointer{ty3_gen}{ty3_gen_lref}{ty3/fwd/ty3\_gen\_lref.con}
\pointer{ty3_gen}{ty3_gen_bind_abbr}{ty3/fwd/ty3\_gen\_bind.con}
\pointer{ty3_gen}{ty3_gen_bind_abst}{ty3/fwd/ty3\_gen\_bind.con}
\pointer{ty3_gen}{ty3_gen_appl}{ty3/fwd/ty3\_gen\_appl.con}
\pointer{ty3_gen}{ty3_gen_cast}{ty3/fwd/ty3\_gen\_cast.con}

\pointer{ty3_props}{ty3_lift}{ty3/props/ty3\_lift.con}
\pointer{ty3_props}{ty3_correct}{ty3/props/ty3\_correct.con}
\pointer{ty3_props}{ty3_unique}{ty3/props/ty3\_unique.con}
\pointer{ty3_props}{ty3_fsubst0}{ty3/fsubst0/ty3\_fsubst0.con}
\pointer{ty3_props}{ty3_subst0}{ty3/fsubst0/ty3\_subst0.con}
\pointer{ty3_props}{ty3_csubst0}{ty3/fsubst0/ty3\_csubst0.con}
\pointer{ty3_props}{csubt_ty3}{csubt/ty3/csubt\_ty3.con}
\pointer{ty3_props}{ty3_typecheck}{ty3/props/ty3\_typecheck.con}

\pointer{ty3_sred}{ty3_sred_wcpr0_pr0}{ty3/pr3/ty3\_sred\_wcpr0\_pr0.con}
\pointer{ty3_sred}{ty3_sred_pr3}{ty3/pr3/ty3\_sred\_pr3.con}
\pointer{ty3_sred}{ty3_gen_lift}{ty3/pr3\_props/ty3\_gen\_lift.con}
\pointer{ty3_sred}{ty3_tred}{ty3/pr3\_props/ty3\_tred.con}
\pointer{ty3_sred}{ty3_sconv_pc3}{ty3/pr3\_props/ty3\_sconv\_pc3.con}
\pointer{ty3_sred}{ty3_sconv}{ty3/pr3\_props/ty3\_sconv.con}

\pointer{ty3_arity_props}{ty3_arity}{ty3/arity/ty3\_arity.con}
\pointer{ty3_arity_props}{ty3_sn3}{ty3/arity\_props/ty3\_sn3.con}
\pointer{ty3_arity_props}{csubt_csuba}{csubt/csuba/csubt\_csuba.com}
\pointer{ty3_arity_props}{ty3_predicative}{ty3/arity\_props/ty3\_predicative.con}
\pointer{ty3_arity_props}{ty3_repellent}{ty3/arity\_props/ty3\_repellent.con}
\pointer{ty3_arity_props}{ty3_acyclic}{ty3/arity\_props/ty3\_acyclic.con}

\pointer{ty3_dec}{pc3_dec}{pc3/dec/pc3\_dec.con}
\pointer{ty3_dec}{ty3_inference}{ty3/dec/ty3\_inference.con}

\pointer{sty_props}{ty3_sty0}{ty3/sty0/ty3\_sty0.con}
\pointer{sty_props}{sty1_env}{sty1/env/sty1\_env.con}

\pointer{levels_ex0}{leqz_leq}{ex0/props/leqz\_leq.con}
\pointer{levels_ex0}{leq_leqz}{ex0/props/leq\_leqz.con}

\pointer{ty3_ex1}{ex1_arity}{ex1/props/ex1\_arity.con}
\pointer{ty3_ex1}{ex1_ty3}{ex1/props/ex1\_ty3.con}

\pointer{nf2_ex2}{ex2_nf2}{ex2/props/ex2\_nf2.con}
\pointer{nf2_ex2}{ex2_arity}{ex2/props/ex2\_arity.con}

\pointer{void_props}{pc3_head_2_void}{pc3/props/pc3\_head\_2.con}
\pointer{void_props}{sc3_void}{sc3/props/sc3\_bind.con}
\pointer{void_props}{ty3_gen_void}{ty3/fwd/ty3\_gen\_bind.con}

\pointer{wf3_props}{wf3_total}{wf3/props/wf3\_total.con}
\pointer{wf3_props}{wf3_ty3_conf}{wf3/ty3/wf3\_ty3\_conf.con}
\pointer{wf3_props}{wf3_ty3}{wf3/props/wf3\_ty3.con}

\end{enumerate}

%% file: thanks.tex
The author would like to thank professor A. Asperti and the whole \textsc{helm}
working group for many valuable discussions on the matter of this text.
The author is also grateful to professor F. Kamareddine, whose papers greatly
influenced this work both theoretically and technically, and to professor
R.P. Nederpelt.